\documentclass[
reprint,
amsmath,
amssymb,
aps,
pra,
10pt,
]{revtex4-2}

\usepackage{graphicx}
\usepackage{dcolumn}
\usepackage{xcolor}
\usepackage{bm}

\begin{document}

\preprint{APS/123-QED}

\title{Exploiting SU($N$) dynamical symmetry for rovibronic stabilization of a weakly bound diatomic molecule}

\author{Diego F. Uribe}
\affiliation{Departamento de Química, Universidad del Valle, A.A. 25360, Cali, Colombia.}

\author{Mateo Londoño}
\affiliation{Departamento de Física, Universidad del Valle, A.A. 25360, Cali, Colombia.}
\altaffiliation{Present address: Department of Physics and Astronomy and Institute for Advanced Computational Science, Stony Brook University, Stony Brook, New York 11794, USA.}

\author{Julio C. Arce}
\email{julio.arce@correounivalle.edu.co}
\affiliation{ Departamento de Química, Universidad del Valle, A.A. 25360, Cali, Colombia.}



\date{\today}

\begin{abstract}

\noindent We propose a multilevel scheme to coherently transfer the population of a diatomic molecule from a rovibrational level to a target rovibrational level of the same electronic state or another.
It involves a linear chain of $N$ rovibrational levels alternating between the initial electronic state and a second electronic state, conveniently selected according to the dipole couplings between consecutive levels.
A set of $N-1$ simultaneous weak laser $\pi$ pulses, with simple analytical shapes, each in resonance between two neighbors of the chain, transfers the population from the initial rovibronic state gradually and consecutively through the chain, until at the end of the process it resides in the target rovibronic state, as in a kind of ping-pong game between the two electronic states.
Using the partial-wave expansion of the molecular wave function, vibrational bases within the $J$ manifolds of each electronic state, and the rotating-wave approximation (RWA), we map the radial Hamiltonian to the one of a spin $s=(N-1)/2$ under a static magnetic field, providing an analytical formula for the populations of the linked states.
As an illustration, we apply the scheme to the stabilization into the absolute ground state of a KRb molecule initially in the high-lying $\upsilon=75$, $J=6$ level of the ground electronic state $X^{1}\Sigma^{+}$. With a chain of seven rovibronic states, three of them belonging to the excited $A^{1}\Sigma^{+}$ electronic state, and pulses of 0.4 ns of duration, the population is fully transferred into the target state in about 1 ns.
In addition, we perform a numerical simulation of the process, taking into account the entire rovibrational space of both electronic states and without invoking the RWA, finding a good agreement with the results of the pseudospin model.
\end{abstract}

\maketitle

\section{\label{sec:Intro}Introduction}

The formation and application of gaseous samples of diatomic molecules at temperatures below 1 K is a research program of continuing interest \cite{Weiner1999,Carr2009,Bohn2017,Dulieu2018,perezriosbook}.
For example, they constitute a novel platform for the quantum simulation of strongly correlated many-body systems with long-range and anisotropic interactions, such as the dipole-dipole interaction \cite{Blackmore2019, Schuster2021, Hughes2023}, the study of state-to-state chemical reactions \cite{Ni2021, Ni2022}, and the creation of new quantum states of matter \cite{Deng2023, Duda2023}.

Two main classes of strategies to achieve that goal have been investigated. Direct methods begin with a sample of molecules at ambient temperature followed by a cooling scheme. Hence, they are very challenging, as they must deal with wide vibrational, rotational, and translational population distributions \cite{Parazzoli2009, Krems.2009, Manai2012, Bigelow2012, Lemeshko2013,Hamamda2015}. On the other hand, indirect methods start by cooling a sample of atoms, followed by an association mechanism, like magnetoassociation (MA) \cite{Marcelis2008,Carr2009,Thalhammer2009, Chin2010,Wacker2015} or photoassociation (PA) \cite{Ulmanis2012,Molano2019,Eisele2020}, which produces translationally cold, albeit vibrationally and rotationally excited, diatoms bonded in a single electronic state \cite{Stevenson2016, Aikawa2010}. Subsequently, a stabilization scheme that drives the molecules into low-lying rovibrational levels of the ground electronic state, hopefully including the absolute ground state, must be applied.

Laser stabilization schemes employ long pulses \cite{Koch2012}, short pulses \cite{Salzmann2007, Wang2017}, chirped pulses \cite{Carini2016}, anti-intuitive pairs of pulses, such as STIRAP \cite{Ni2008}, or the selection of a suitable excited state to enhance the spontaneous decay to a deep bound rovibronic level \cite{Stevenson2016}. Theoretical explorations of alternative schemes include quantum coherent control methods \cite{Koch2012, DeLima2017}, quantum optimal control algorithms \cite{DeLima2015,Ndong2010,Guerrero2018,Londono2023}, or the use of highly nonresonant radiation to endow a collision state with shape-resonance character \cite{Gonzalez-Ferez2012, Crubellier2015}.

Here, we propose a stabilization scheme that drives a molecule, either polar or nonpolar, from an initial into a target rovibrational level, in general belonging to a different electronic state, with very high efficiency.
In Sec. \ref{sec:Scheme}, within the context of a prototype KRb molecule formed in a high-lying rovibrational level of the ground electronic state, we explain the scheme, which entails a chain of $N$ rovibrational levels alternating between the initial electronic state and a second electronic state, with $N-1$ simultaneous weak laser pulses transferring the population along the chain in a manner that resembles a ping-pong game between the two electronic states.
In Sec. \ref{sec:Method}, employing the partial-wave expansion of the molecular wavefunction, neglecting the dipole coupling between rovibrational levels of the same electronic state, introducing vibrational bases within the $J$ manifolds of each electronic state, and invoking the rotating-wave approximation (RWA), we map the radial Hamiltonian matrix into the matrix representation of a spin with quantum number $s=(N-1)/2$ under a static magnetic field, producing an analytical formula for the populations of the selected rovibronic states. In addition, we explain our choice of analytical pulse shapes and the methodologies for the numerical calculation of the rovibrational levels within each electronic state and the integration of the coupled radial time-dependent equations, taking into account the full rovibrational structure of both electronic states and without using the RWA.
In Sec. \ref{sec:results} we apply the pseudospin representation to the prototype KRb molecule and discuss how potential complications, considering the complexity of its rovibronic structure, actually do not hamper its validity. Then, we present the results of the analytical formula and the numerical simulations, which are found to agree very well.
Finally, in Sec. \ref{sec:conclusions} we state the conclusions of our results and present some perspectives for future work.

\section{\label{sec:Scheme}Ping-pong Scheme}

\begin{figure}
    \centering
    \scalebox{0.12}{\includegraphics{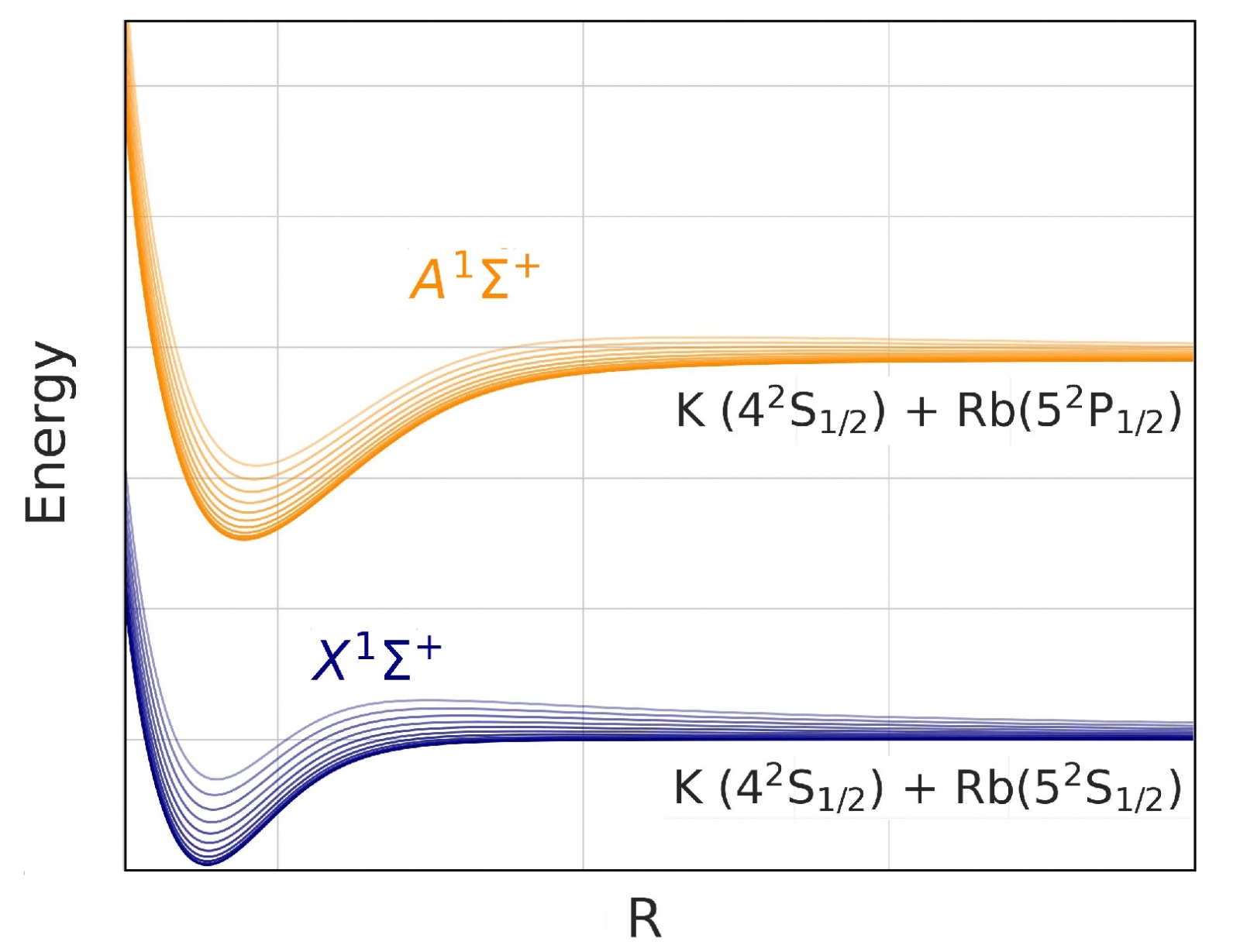}}
    \caption{Schematic representation (not to scale) of the $J=0$ and some $J>0$ potential energy curves associated with the ground and excited electronic states of KRb.}
    \label{fig:PECs_graph}
\end{figure}

Our rovibronic stabilization scheme consists of an adaptation of the Cook-Shore multilevel coherent excitation chain \cite{Shore2011}. For its implementation, we chose as prototype a KRb molecule, since it has a complicated rovibronic structure, in particular, a strong spin-orbit (SO) coupling between the two low-lying excited electronic states $A^1\Sigma^+$ and $b^3\Pi$, and a high density of rovibrational levels near the dissociation threshold of the ground electronic state $X^1\Sigma^+$, providing a stringent test of the validity of the scheme. Figure \ref{fig:PECs_graph} schematically illustrates the potential energy curves (PECs) for $J=0$ and several $J>0$ of the $X^1\Sigma^+$ and $A^1\Sigma^+$ states.

\begin{figure}
    \centering
    \scalebox{0.12}{\includegraphics{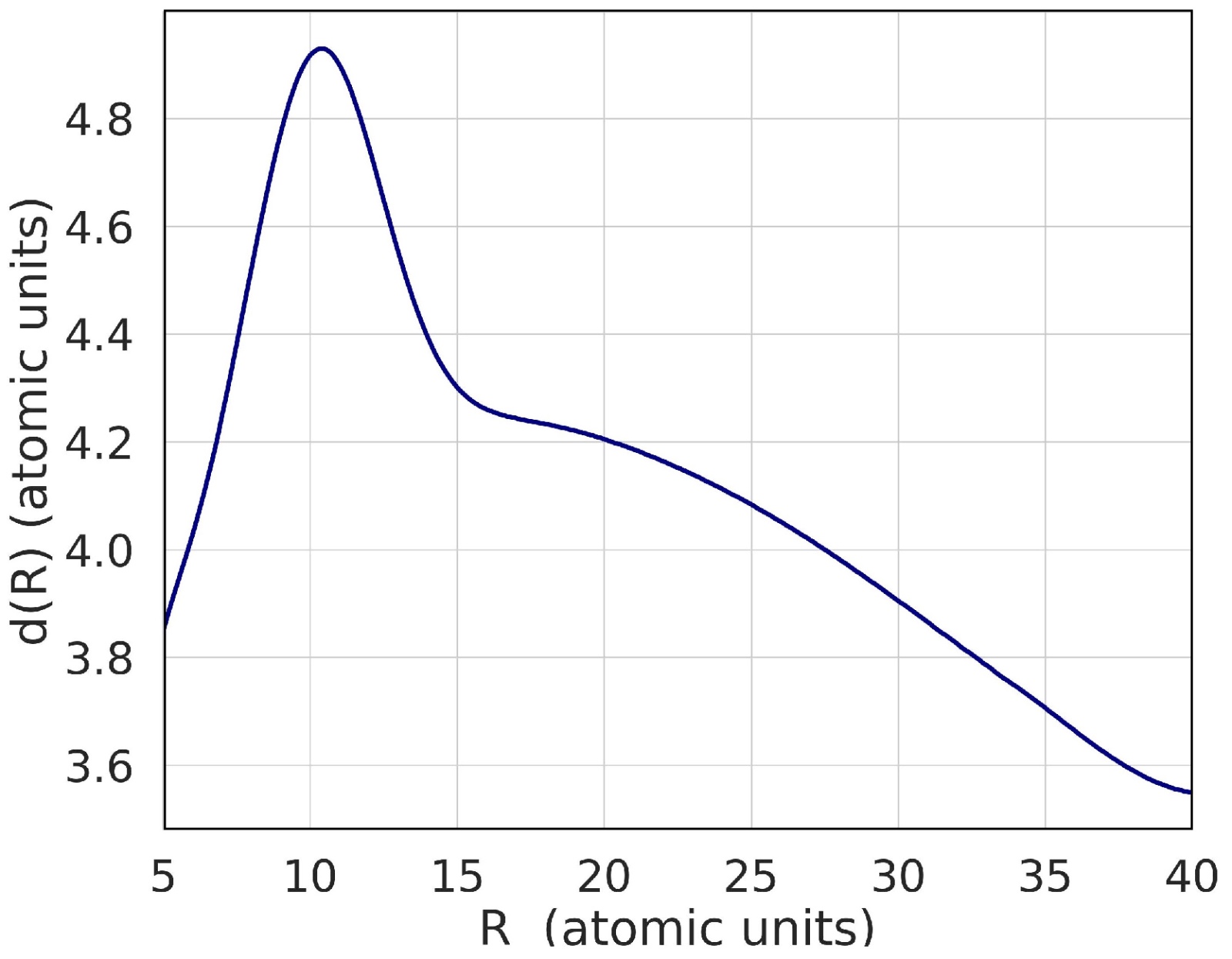}}
    \caption{Transition dipole moment function between the $X^{1}\Sigma^{+}$ and $A^{1}\Sigma^{+}$ states of KRb.}
    \label{fig:xadipole}
\end{figure}

The familiar one-photon E1 selection rule, $\Delta J=\pm 1$, dictates that even-$J$ levels of the $X^1\Sigma^+$ state are coupled to odd-$J$ levels of the $A^1\Sigma^+$ state and viceversa. Although the dynamics of these two types of transitions develop simultaneously, they are independent. Since we target the $J=0$ vibrational manifold of the ground electronic state, we focus on the first type of transition. Figure \ref{fig:xadipole} displays the transition dipole moment between the $X^1\Sigma^+$ and $A^1\Sigma^+$ states. In principle, the permanent dipole moments of both electronic states can allow intracurve rovibrational transitions. However, we do not take them into account, because the frequencies of the applied field are tuned in resonance with rovibronic transitions, whose energy differences are much higher than the ones of rovibrational transitions. Thus, our scheme also applies to nonpolar molecules.

\begin{figure}
    \centering
    \scalebox{0.36}{\includegraphics{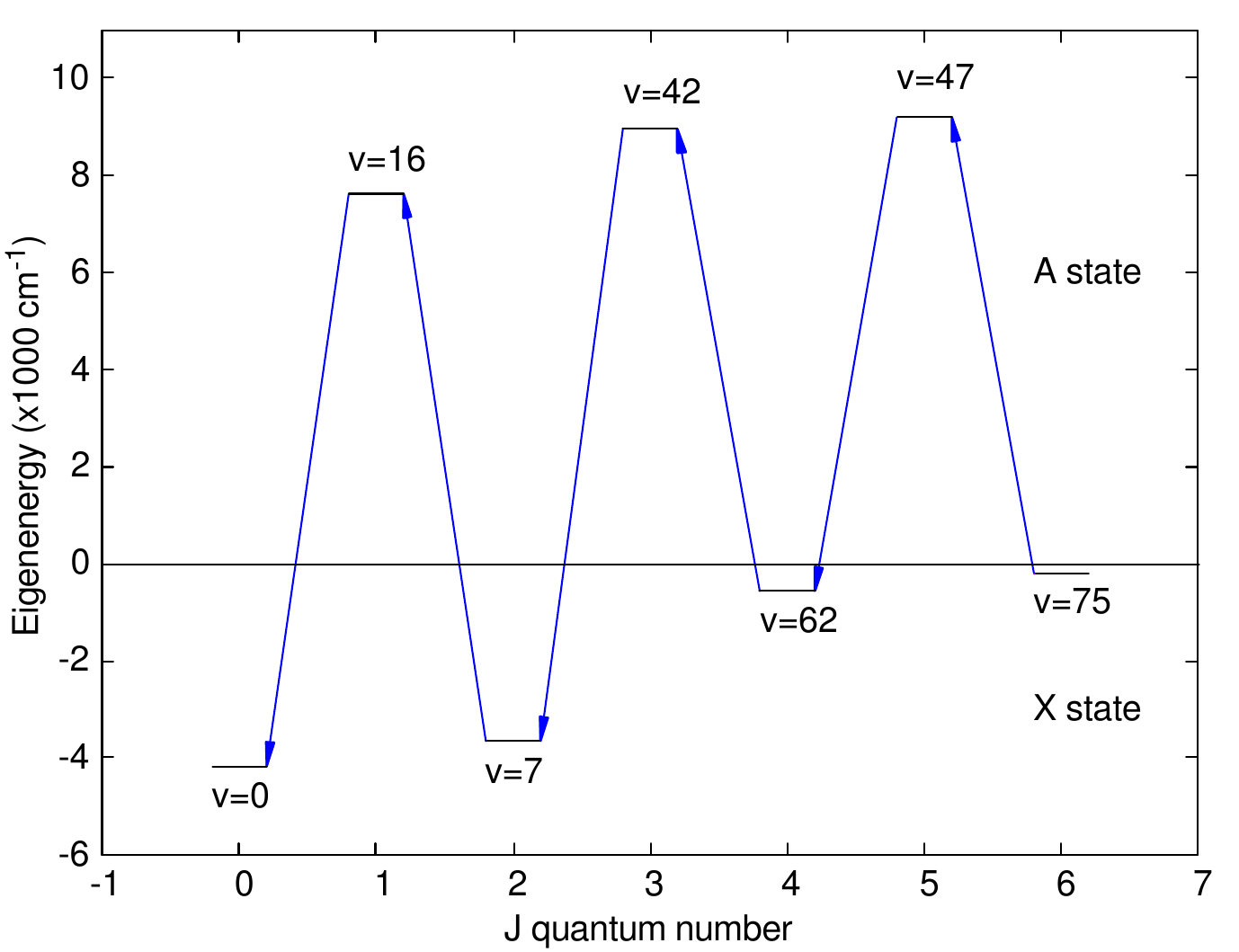}}
    \caption{Stabilization route across the rovibrational spaces of the ground $X^1\Sigma^+$ and excited $A^1\Sigma^+$ states of KRb. Below and above the horizontal line the vibrational quantum numbers belong to the ground and excited electronic states, respectively.}
    \label{fig:route}
\end{figure}

We assume that the molecule has been formed in a highly-excited rovibrational level of the $X^1\Sigma^+$ state, as occurs in one-step PA \cite{Molano2019}. A laser pulse brings this level into resonance with a rovibrational level of the $A^1\Sigma^+$ state. A second laser pulse brings the latter into resonance with a level of the $X^1\Sigma^+$ state, lying lower than the initial level. Other pulses drive transitions up and down, with each level in the $X^1\Sigma^+$ state lying lower than the previous one, until the absolute ground state, $|X^1\Sigma^+,\upsilon=0,J=0\rangle$ is reached, hopefully with a high probability. We use pulses weak and long enough so that their bandwidths do not include any other Bohr frequencies than the desired ones. The resulting mechanism is a kind of ping-pong game between the two electronic states. Figure \ref{fig:route} shows an example of a linkage pattern.

The linked states must be selected so that the squares of the dipole matrix elements (DMEs)

\begin{equation}
\label{eq:DME}    
d_{\upsilon,J;\upsilon^{\prime},J^{\prime}} = \langle{X^1\Sigma^+,\upsilon,J} \big|d(R)\big|{A^1\Sigma^+,\upsilon^{\prime},J^{\prime}}\rangle
\end{equation}

\noindent have relatively large values.

\section{\label{sec:Method}Methodology}

\subsection{\label{sec:modelH}Model rovibronic Hamiltonian}

The ground, $X$, and excited, $A$, states of the diatomic molecule are coupled by a radiation field. Within the two-state and semiclassical electric dipole (E1) approximations, the Hamiltonian takes the form

\begin{align}\label{eq:ham-matrix}
&\hat{\mathbf{H}}(t) = \nonumber\\
&\left(
\begin{matrix}
\hat{H}_X-\hat{\epsilon}\cdot\hat{R}d_{X}(R)E(t) & -\hat{\epsilon}\cdot\hat{R} d(R)E(t) \\
-\hat{\epsilon}\cdot\hat{R} d(R)E(t)  & \hat{H}_A-\hat{\epsilon}\cdot\hat{R} d_{A}(R)E(t) \\
\end{matrix}
\right),
\end{align}

\noindent where $\hat R$ is a unitary vector in the direction of the internuclear vector $\vec R=\hat R R$, $\hat{H}_X$ and $\hat{H}_A$ are the field-free Hamiltonians of the electronic states, $d_X(R)$ and $d_A(R)$ are the permanent E1 moments, $d(R)$ is the transition E1 moment, and $\hat\epsilon$ and $E(t)$ are the polarization vector and amplitude of the electric field.
The time-dependent Schr\"odinger equation (TDSE) reads (here and henceforth we employ atomic units)

\begin{equation}\label{eq:TDSE-matrix}
i\dot{\mathbf{\Psi}}(\vec{R},t)=\hat{\mathbf{H}}(t)\mathbf{\Psi}(\vec{R},t), 
\end{equation}

\noindent with the two-component wave function

\begin{equation}\label{eq:Psi-column}
\mathbf{\Psi}(\vec{R},t)=\left(
\begin{matrix}
\Psi_X(\vec{R},t) \\
\Psi_A(\vec{R},t) \\
\end{matrix}
\right).
\end{equation}

In the nuclear center of mass, we expand each component in partial waves,

\begin{equation}\label{eq:pw-expansion}
\Psi_e(R,\theta,\phi,t)=\sum_J \sum_M F_J^e(R,t)Y_J^M(\theta,\phi),
\end{equation}

\noindent with $e=X,A$. Hence, the radial TDSE becomes

\begin{equation}\label{eq:TDSE-pw}
i{\mathbf{\dot{F}}}(R,t)= \mathbf{\hat{H}_R}(t){\mathbf{F}}(R,t).
\end{equation}

\noindent where $\mathbf F(R,t) $ is the multi-$J$-component radial function and $\mathbf{\hat{H}_R}(t)$ is the radial Hamiltonian. We assume that the field is $z$-polarized and the initial state of the molecule has $M=0$. For the sake of illustration, let us suppose, for the time being, that the $X$ and $A$ electronic states had only two rotational levels, $J=0,1$. Then, taking into account the selection rules $\Delta J=\pm 1$ and $\Delta M=0$,

\begin{equation}\label{eq:F-column}
\mathbf{F}(R,t)=\left(
\begin{matrix}
F_0^X(R,t) \\
F_1^X(R,t) \\
F_0^A(R,t) \\
F_1^A(R,t)
\end{matrix}
\right),
\end{equation}

\begin{align}\label{eq:ham-pw1}
&\mathbf{\hat{H}_R}(t) = \nonumber\\
&\left(
\begin{matrix}
 \hat{T} + \hat{V}_{0}^{X}  &  -a_{0}d_{X}E(t)  &  0  &  -a_{0}dE(t)  \\
 -b_{1}d_{X}E(t)    &  \hat{T} + \hat{V}_{1}^{X} & -b_{1}dE(t) & 0 \\
 0 & -a_{0}dE(t)    & \hat{T} + \hat{V}_{0}^{A} & -a_{0}d_{A}E(t) \\
 -b_{1}dE(t) & 0 & -b_{1}d_{A}E(t)    & \hat{T} + \hat{V}_{1}^{A} \\
\end{matrix}
\right),
\end{align}

\noindent where

\begin{equation}
\hat T=- \frac{1}{2\mu} \frac{d^{2}}{d R^{2}},
\end{equation}

\begin{equation}\label{eq:V_eff}
V^e_J(R) = \frac{J(J+1)}{2\mu R^{2}} + U^e(R),
\end{equation}

\noindent with $\mu$ and $U^e(R)$ being the reduced mass and the PEC of the $e$ electronic state of the molecule, and $a_J$ and $b_J$ being the angular matrix elements

\begin{align}\label{eq:ang-fac}\nonumber
    a_{J}&=\langle Y_J^M|\cos\theta|Y_{J+1}^M\rangle=\left[ \frac{(J+1)^{2}}{(2J+3)(2J+1)} \right]^{1/2},  \\
    b_{J}&=\langle Y_J^M|\cos\theta|Y_{J-1}^M\rangle=\left[ \frac{J^{2}}{(2J+1)(2J-1)} \right]^{1/2}.
\end{align}

As justified in Sec. \ref{sec:Scheme}, we neglect intracurve rovibrational transitions. In addition, by taking into account that $b_J=a_{J-1}$ and rearranging the matrix of Eq. (\ref{eq:ham-pw1}) we obtain for the radial Hamiltonian

\begin{align}\label{eq:ham-pw2}
&\mathbf{\hat{H}_R}(t) = \nonumber\\
&\left(
\begin{matrix}
 \hat{T} + \hat{V}_{0}^{X}  &  -a_{0}dE(t)  &  0  &  0\\
 -a_{0}dE(t) &  \hat{T} + \hat{V}_{1}^{A} & 0 & 0 \\
 0 & 0 & \hat{T} + \hat{V}_{0}^{A} & -a_{0}dE(t) \\
 0 & 0 & -a_{0}dE(t) & \hat{T} + \hat{V}_{1}^{X} \\
\end{matrix}
\right),
\end{align}

\noindent which has a block-diagonal structure, where even $J$ values of the $X$ state are coupled to odd $J$ values of the $A$ state, and viceversa. Clearly, the dynamics of these two types of transitions develop independently.

Since our target is the absolute ground state, $|X^1\Sigma^+,\upsilon=0,J=0\rangle$, now we can appreciate that the general structures of the radial multicomponent function and Hamiltonian are


\begin{equation}\label{eq:F-column2}
\mathbf{F}(R)=\left(
\begin{matrix}
F_0^X(R,t) \\
F_1^A(R,t) \\
F_2^X(R,t) \\
F_3^A(R,t) \\
\vdots
\end{matrix}
\right),
\end{equation}

\begin{align}\label{eq:ham-pw3}
&\hat{\mathbf{H}}(t) = \nonumber\\
&\left(
\begin{matrix}
 \hat{T} + \hat{V}_{0}^{X}  &  -a_{0}dE(t)  &  0  &  0 & \cdots \\
 -a_{0}dE(t)    &  \hat{T} + \hat{V}_{1}^{A} & -a_{1}dE(t) & 0 & \cdots \\
 0 & -a_{1}dE(t)    & \hat{T} + \hat{V}_{2}^{X} & -a_{2}dE(t) & \cdots \\
 0 & 0 & -a_{2}dE(t)    & \hat{T} + \hat{V}_{3}^{A} & \cdots \\
 \vdots & \vdots & \vdots & \vdots & \ddots
\end{matrix}
\right).
\end{align}

\subsection{\label{sec:RWA}RWA in the rovibronic space}

We represent the radial function of the $J$ manifold within the $e$ electronic state in the vibrational eigenbasis \{$\varphi_{\upsilon,J}^{e}(R)$\} as

\begin{align}\label{eq:rv-expansion}
{F_{J}^{e}(R,t)} = \sum_{\upsilon} c_{\upsilon,J}^{e}(t) e^{i\omega_{\upsilon,J}^e t} \varphi_{\upsilon,J}^{e}(R),
\end{align}


\noindent where $\omega_{\upsilon,J}^e$ are the eigenfrequencies of the rovibronic states.
By plugging these expansions into Eq. (\ref{eq:TDSE-pw}) with the Hamiltonian (\ref{eq:ham-pw3}), we get the system of equations for the time-dependent coefficients

\begin{align}
\label{eq:coeff}
\dot{c}_{\upsilon} = i a E(t) \sum_{\upsilon'} d_{\upsilon \upsilon'} c_{\upsilon'} & e^{i\omega_{\upsilon \upsilon'}t} \nonumber\\
+& i b E(t) \sum_{\upsilon''} d_{\upsilon \upsilon''} c_{\upsilon ''} e^{i\omega_{\upsilon \upsilon''}t}.
\end{align}

\noindent Here, to ease the notation, we have dropped the $e$ and $J$ indexes, without risk of confusion, since now $\upsilon'$ and $\upsilon''$ run over the $J+1$ and $J-1$ manifolds, respectively, which belong to different electronic states. For the $J+1$ manifold, $\omega_{\upsilon\upsilon'}\equiv \omega_{\upsilon}-\omega_{\upsilon'}$ and $d_{\upsilon\upsilon'}$ is given by Eq. (\ref{eq:DME}), and analogously for the $J-1$ manifold. $a$ and $b$ are given by Eqs. (\ref{eq:ang-fac}).

Our ping-pong stabilization scheme involves a chain of $N$ levels, each one effectively coupled (linked) only to its nearest neighbors (the chain-end levels have only one nearest neighbor, while the rest have two). Thus, the field is a superposition of $N-1$ laser pulses,

\begin{equation}\label{eq:laser}
E(t)=\sum_{k=1}^{N-1} E^k(t)=\sum_{k=1}^{N-1}\varepsilon^k(t)\cos(\omega^k t),
\end{equation}

\noindent where $\varepsilon^k(t)$ and $\omega^k$ are the temporal envelope and carrier frequency of the $k$-th pulse, which is in resonance only with two levels of the chain and is long enough so that its bandwidth includes, by far, mainly the desired Bohr frequency. For the $J+1$ manifold, the detunings and resonance Rabi frequencies are

\begin{align}\label{eq:omega_kvv}
&\Delta^{k}_{\upsilon \upsilon'}=\omega_{\upsilon\upsilon'}-\omega^{k},
&\Omega^{k}_{\upsilon\upsilon'}(t)={a\varepsilon^{k}(t)d_{\upsilon\upsilon'}}/{{2}},
\end{align}

\noindent respectively, and analogously for the $J-1$ manifold.
All pulses are weak enough so as to induce Rabi frequencies low enough such that $\Delta \gg \Omega$ for all unlinked levels.

Under these conditions, we can apply the rotating-wave approximation (RWA) \cite{Shore2011}, which entails the neglect of the rapidly oscillating antiresonant terms $e^{i(\omega^{k}+\omega_{\upsilon\upsilon'})t}$ and $e^{i(\omega^{k}+\omega_{\upsilon\upsilon''})t}$, producing

\begin{equation}
\label{eq:rabiRWA}
\dot{c}_{\upsilon} = i \sum_{k}\sum_{\upsilon'} \Omega^{k}_{\upsilon \upsilon'} c_{\upsilon'} e^{i\Delta^{k}_{\upsilon\upsilon'}t} +i \sum_{k}\sum_{\upsilon''} \Omega^{k}_{\upsilon\upsilon''} c_{\upsilon''} e^{i\Delta^{k}_{\upsilon\upsilon''}t}.
\end{equation}

\noindent Moreover, since $e^{i\Delta^{k}_{\upsilon\upsilon'}t}=1$ when the resonance condition $\omega^{k}-\omega_{\upsilon\upsilon'}=0$ is met, we obtain the system

\begin{equation}\label{eq:rabi2}
\dot{c}_{\upsilon} = i \Omega^{k'}_{\upsilon\upsilon'} c_{\upsilon'} + i \Omega^{k''}_{\upsilon\upsilon''} c_{\upsilon''},
\end{equation}

\noindent where $k'$ and $k''$ are the pulses that drive the $\upsilon,J\rightarrow\upsilon',J+1$ and $\upsilon,J\rightarrow\upsilon'',J-1$ transitions, respectively, between linked levels. The system (\ref{eq:rabiRWA}) can now be written as

\begin{equation}\label{eq:TDSE-expansion}
\mathbf{\dot{C}}(t)= i\mathbf{\hat{W}}(t){\mathbf{C}(t)},
\end{equation}

\noindent with $\mathbf{C}(t)$ being the vector of coefficients and $\mathbf{\hat{W}}(t)$ being the RWA Hamiltonian matrix

\begin{equation}\label{eq:wmatrix}
\hat{\mathbf{W}}(t) = 
\left(
\begin{matrix}
 0  &  \Omega^{k'}  &  0  &  0 & \cdots \\
 \Omega^{k'}  &  0 & \Omega^{k''} & 0 & \cdots \\
 0 & \Omega^{k''}  & 0 & \Omega^{k'''} & \cdots \\
 0 & 0 & \Omega^{k'''} & 0 & \cdots \\
 \vdots & \vdots & \vdots & \vdots & \ddots
\end{matrix}
\right).
\end{equation}

\subsection{\label{sec:PSR}Pseudospin representation}

The structure of the matrix (\ref{eq:wmatrix}) permits mapping
our model to that of a spin $s=(N-1)/2$ under the influence of a magnetic field, such that the $2s+1=N$ magnetic substates represent the linked molecular rovibronic states.
This mapping can be accomplished by setting the Rabi frequencies of Eq. (\ref{eq:omega_kvv}) to

\begin{equation}\label{eq:omega_k}
    \Omega^{k}_{\upsilon\upsilon'}=\Omega_{0}(t)\sqrt{k(N-k)},
\end{equation}

\noindent with $k=1,2,3,\dots,N-1$ and $\Omega_{0}(t)$ being a common time-dependent profile \cite{Shore2011}. Thus, in general, the RWA Hamiltonian (\ref{eq:wmatrix}) can be expressed as

\begin{equation}\label{eq:wps}
    \hat{\mathbf{W}}(t) = \text{Re}\{\Omega_{0}(t)\}\hat{\mathbf{S}}_{x} + \text{Im}\{\Omega_{0}(t)\}\hat{\mathbf{S}}_{y} + \Delta_{0} \hat{\mathbf{S}}_{z} = {\mathbf{\Gamma}} \cdot \hat{\mathbf{S}},
\end{equation}

\noindent where $\mathbf{\Gamma}$ is the magnetic field vector, whose Cartesian components are $\text{Re}\{\Omega_{0}(t)\}, \text{Im}\{\Omega_{0}(t)\}, \Delta_{0}$, with $\Delta_{0}$ being the cumulative detuning, and $\hat{\mathbf{S}}$ is the matrix representative of the spin $s$, whose Cartesian components are $\hat{\mathbf{S}}_{x}, \hat{\mathbf{S}}_{y}, \hat{\mathbf{S}}_{z}$. In our case, the Rabi frequencies are real and $\Delta_{0}=0$. Hence, the RWA Hamiltonian matrix becomes

\begin{equation}\label{eq:H_RWA}
\hat{\mathbf{W}}(t)=\Omega_{0}(t)\hat{\mathbf{S}}_{x}.
\end{equation}

The population of any rovibronic state in the chain, when the initial population resides in one of the chain-end states, turns out to be given by \cite{Shore2011, Hioe1987}

\begin{equation}\label{eq:psm}
\big|{c}_{m}(t)\big|^{2} = \binom{2s}{s-m} \big|\cos (\Omega_{0}t)\big|^{2(s-m)} \big|\sin (\Omega_{0}t)\big|^{2(s+m)},
\end{equation}

\noindent where the prefactor is a binomial coefficient. In this representation, the magnetic substates $m=-s,\dots,+s$ are related to the vibronic states $i=1,2,3,\dots,N$ as per $m=i-s-1$.

\subsection{\label{sec:num}Numerical simulation}

The TDSE (\ref{eq:TDSE-pw}) with the Hamiltonian (\ref{eq:ham-pw3}) consists of the set of coupled equations for the radial functions

\begin{align}
\label{eq:coupledeq}
i{\dot{F}_{J}^{e}}(R,t) = \left[ -\frac{1}{2\mu} \frac{d^{2}}{d R^{2}} + \hat{V}^{e}_{J}(R) \right] &{F_{J}^{e}}(R,t) \nonumber\\
-d(R)E(t)a_{J}{F_{J+1}^{e'}}(R,t) \nonumber\\
-d(R)E(t)b_{J}{F_{J-1}^{e'}}(R,t),
\end{align}

\noindent where the applied field $E(t)$ is given by Eq. (\ref{eq:laser}) and for $e=X$ or $e=A$ only even or odd $J$ values are considered, respectively. We integrated these equations numerically.

We evaluated the level populations

\begin{equation}\label{eq:pop}
p^e_{\upsilon,J}(t)=|\langle\varphi^e_{\upsilon,J}|F^e_{J}(t)\rangle|^2
\end{equation}

\noindent by numerical quadrature.

Since Eqs. (\ref{eq:coupledeq}) do not involve the RWA or any particular level linkage, a comparison of these numerical populations with the analytical ones provided by Eq. (\ref{eq:psm}) will provide a stringent test of the validity of the approximations introduced in Sec. \ref{sec:RWA}.

\subsection{Computational details}

We took the $X^{1}\Sigma^{+}$ and $A^{1}\Sigma^{+}$ PECs and their transition dipole moment function $d(R)$ (see Fig. \ref{fig:xadipole}) from Ref. \cite{Borsalino2014}. We calculated the vibrational eigenenergies and eigenfunctions within each rotational manifold of the $^{39}$K$^{87}$Rb isotopologue by numerical integration of the time-independent nuclear Schrödinger equation

\begin{align}
\label{eq:1dnuclear}
    \left[ - \frac{1}{2\mu} \left( \frac{d^{2}}{d R^{2}} \right) + V^e_J(R) \right]\varphi^e_{\upsilon,J}(R) = E^{e}_{\upsilon,J} \varphi^e_{\upsilon,J}(R)
\end{align}

\noindent using the Colbert-Miller discrete-variable representation method \cite{Colbert1992}. We obtained converged results with a grid of 35 bohr and 7001 grid points.

We represented the radial functions of Eq. (\ref{eq:coupledeq}) on a spatial grid of 20 bohr and 2001 points. We approximated the short-time evolution operator using the split-operator method, which is accurate to $\cal{O}$$(\delta t^3)$. We obtained well-converged results by including the $J=0,2,4,6,8,10$ PECs for the $X$ state and the $J=1,3,5,7,9$ PECs for the $A$ state. We adapted the computer code employed in Refs. \cite{Guerrero2018,Londono2023}. 

For the pulses, we chose Gaussian-4 envelopes

\begin{equation}\label{gauss4}
    \varepsilon^{k}(t) = \varepsilon_{0}^{k} e^{-(t-t_{0}^{k})^{4}/\sigma^{4}},
\end{equation}

\noindent with which the Rabi frequencies (\ref{eq:omega_kvv}) become

\begin{equation}\label{eq:omega_kvv2}
\Omega^{k}_{\upsilon\upsilon'}(t)={e^{-(t-t_{0}^{k})^{4}/\sigma^{4}}\frac{a\varepsilon_{0}^{k}d_{\upsilon\upsilon'}}{2}}.
\end{equation}

\noindent Here, the time-dependent factor $e^{-(t-t_{0}^{k})^{4}/\sigma^{4}}$ is $\Omega_0(t)$ and the field strengths $\varepsilon_{0}^{k}$ are defined so that the quantities $a\varepsilon_{0}^{k}d_{\upsilon\upsilon'}/2$ correspond to $\sqrt{k(N-k)}$ in Eq. (\ref{eq:omega_k}).

\section{Results and Discussion}\label{sec:results}

\subsection{SU($N$) Hamiltonian}

\begin{figure}
    \centering
    \scalebox{0.37}{\includegraphics{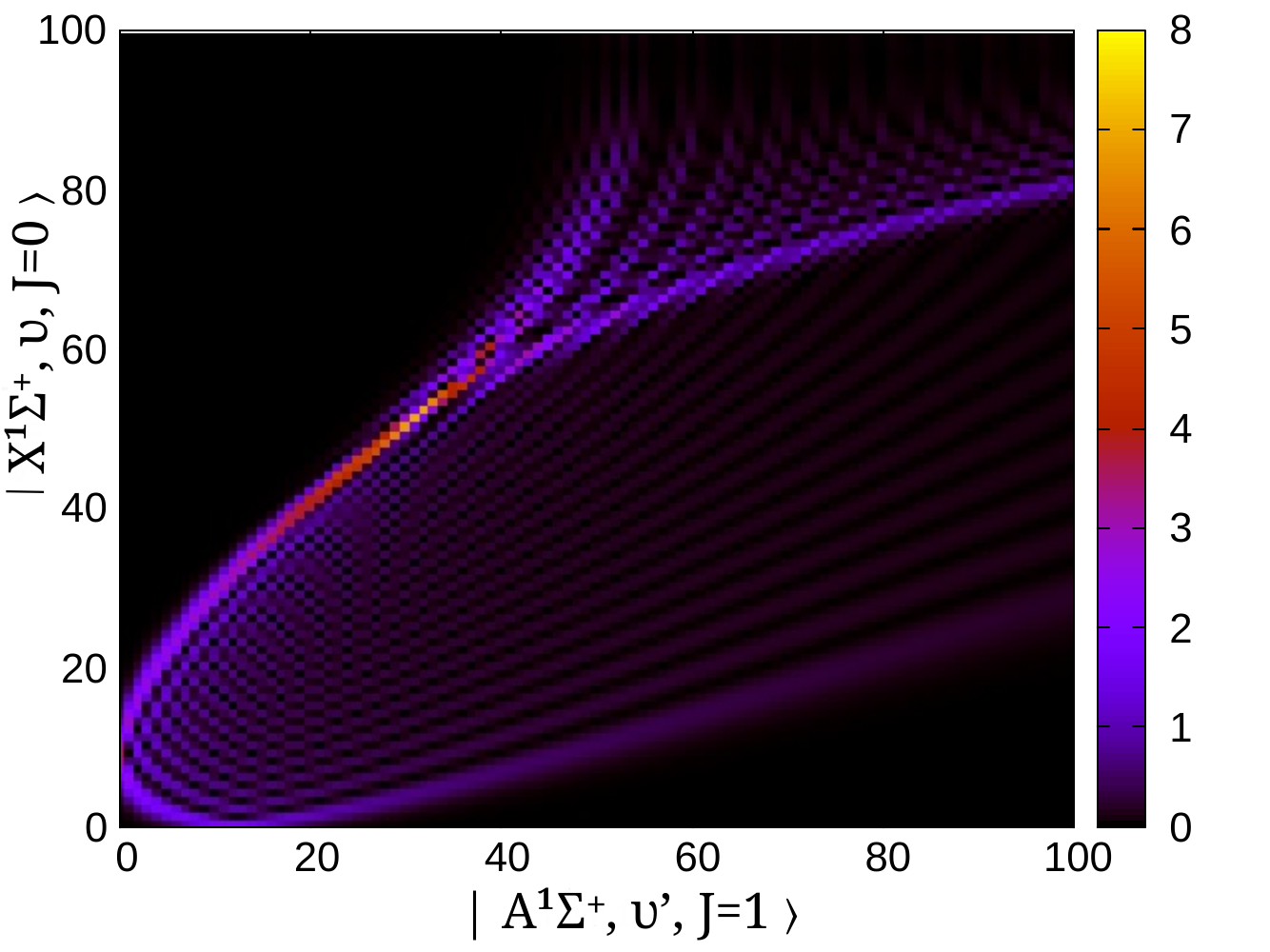}}
    \caption{Squared dipole matrix elements, in atomic units, between the $J=0$ and $J=1$ manifolds of the ground and excited electronic states, respectively.}
    \label{fig:DME}
\end{figure}

Figure \ref{fig:DME} shows the squares of the DMEs between the $J=0$ and $J=1$ manifolds of the electronic states $X^{1}\Sigma^{+}$ and $A^{1}\Sigma^{+}$, respectively; these quantities do not change significantly for other values of $J$. With the aid of coupling maps like this, we selected 7 states to carry out the ping-pong scheme from the initial state $X_{75,6}$ to the absolute ground state $X_{0,0}$. The chain of states is shown in Fig. \ref{fig:route}. 

Now, applying Eq. (\ref{eq:omega_k}) with $N=7$ the RWA Hamiltonian matrix (\ref{eq:wmatrix}) takes the form

\begin{align}\label{7chain}
&\hat{\mathbf{W}}(t) = \nonumber\\
&\Omega_{0}(t)\left(
\begin{matrix}
 0  &  \sqrt{6}  &  0  &  0 & 0 & 0 & 0\\
 \sqrt{6}  &  0 & \sqrt{10} & 0 & 0 & 0 & 0\\
 0 & \sqrt{10}  & 0 & \sqrt{12} & 0 & 0 & 0\\
 0 & 0 & \sqrt{12} & 0 & \sqrt{12} & 0 & 0\\
 0 & 0 & 0 & \sqrt{12} & 0 & \sqrt{10} & 0\\
 0 & 0 & 0 & 0 & \sqrt{10} & 0 & \sqrt{6}\\
 0 & 0 & 0 & 0 & 0 & \sqrt{6} & 0\\
\end{matrix}
\right).
\end{align}

\noindent The elements of this matrix are equal, within a factor of $1/2$, to the elements of the $\hat{\mathbf{S}}_x$ matrix, which are given by $\langle m'|\hat{S}_x|m\rangle=(\delta_{m',m+1}+\delta_{m'+1,m})\frac{1}{2}\sqrt{s(s+1)-m'm}$, with $s=3$. Hence, by redefining $\Omega_0(t)$ so as to include this factor, we obtain the pseudospin representation (\ref{eq:H_RWA}).


\begin{figure}
    \centering
    \scalebox{0.43}{\includegraphics{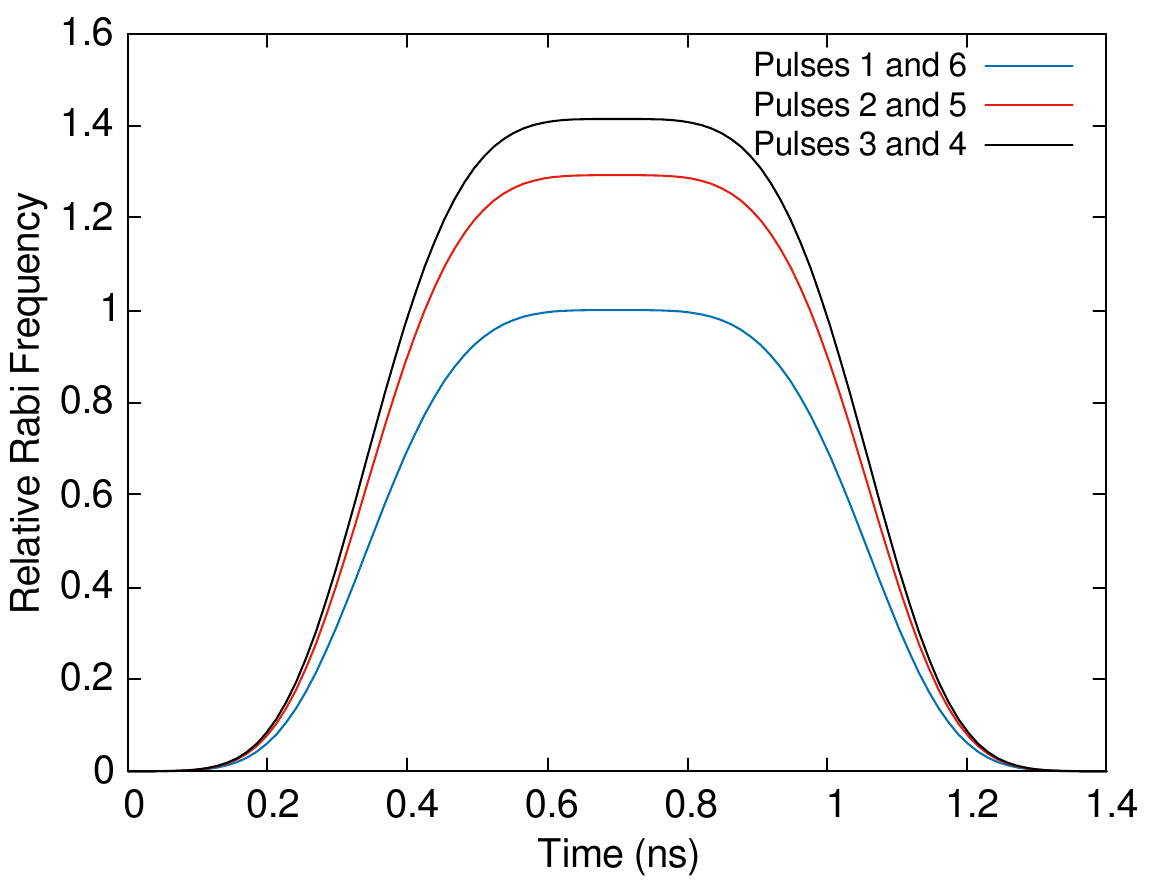}}
    \caption{Time-dependent relative Rabi frequencies employed in the ping-pong scheme.}
    \label{fig:pulses}
\end{figure}

This model Hamiltonian allows us to compute the analytical pulse combination that leads the population through the desired pathway. 
Defining the field strengths as indicated below Eq. (\ref{eq:omega_kvv2}) we obtained $\varepsilon_0^1=1.84\times10^{4}$, $\varepsilon_0^2=1.12\times10^{4}$, $\varepsilon_0^3=1.27\times10^{4}$, $\varepsilon_0^4=7.03\times10^{4}$, $\varepsilon_0^5=4.90\times10^{4}$, and $\varepsilon_0^6=1.18\times10^{4}$, in W/cm$^{2}$.
Figure \ref{fig:pulses} shows the time evolution of the resulting Rabi frequencies (\ref{eq:omega_kvv2}), relative to the one with the lowest maximum, employing pulses with durations of $\sigma=0.387$ ns.
Nevertheless, for the application of these pulses in the complete energy landscape, some potential complications can be envisaged. In the first place, the spin-orbit (SO) coupling induces nonradiative population transfer between the electronic states $A^{1}\Sigma^{+}$ and $b^{3}\Pi$. In addition, radiative decay from the latter into the $a^{3}\Sigma^{+}$ state can occur. However, the time scale of the SO transfer ($\thicksim 1$ ps) \cite{Guerrero2018} is much faster than that of the ping-pong process ($\thicksim 1$ ns), so that there is always population in the $A^{1}\Sigma^{+}$ state to enable the $X-A$ transitions.
Moreover, the role of the triplet $b-a$ spontaneous radiative transitions is limited by two facts. First, the DMEs  for the $b^{3}\Pi$ $\rightarrow$ $a^{3}\Sigma^{+}$ transitions are two orders of magnitude weaker than the ones for the $X^{1}\Sigma^{+}$ $\leftrightarrow$ $A^{1}\Sigma^{+}$ transitions \cite{Guerrero2018}.
Second, the $a^{3}\Sigma^{+}$ well has a depth of 219.6 cm$^{-1}$, which implies that the transitions into this state from the mixed states $A^{1}\Sigma^{+}/b^{3}\Pi$ are in the range 7023 cm$^{-1}$ - 7901 cm$^{-1}$, whereas the lowest energy singlet transition of the ping-pong scheme, $X_{75,6} \rightarrow A_{47,5}$, has an energy of $\approx 9000\ \rm{cm}^{-1}$. Thus, the resonance condition for the triplet transitions can never be met. Hence, it is a good approximation to neglect the effects of the SO coupling and remain in the $X-A$ scheme.


The second complication comes from the high density of levels in the upper regions of the PECs. For example, in Fig. \ref{fig:route} it is seen that the energy of the initial $X_{75,6}$ state is very close to the dissociation threshold, but it turns out that in that energy interval there are $22$ rovibrational levels for $J=6$.
Therefore, even if the bandwidth of a laser pulse is very narrow, it could bring into resonance unwanted states, causing some population to leak out of the chain. 


Additionally to the complications inherent to the ping-pong mechanism, we must assess the role of the spontaneous emission $A^{1}\Sigma^{+}\rightarrow X^{1}\Sigma^{+}$.
The radiative lifetime of the $|A^{1}\Sigma^{+}, \upsilon, J\rangle$ state is given by $\tau_{\upsilon}=\sum_{\upsilon'<\upsilon}A_{\upsilon\upsilon'}^{-1}$, where $A_{\upsilon\upsilon'}=2\omega_{\upsilon\upsilon'}^3 d_{\upsilon\upsilon'}^2/3\epsilon_0 c^3\hbar$ is an Einstein coefficient. The values for the high-lying states $A_{47,5}$ and $A_{42,3}$ are $\tau \sim 10$ ns, which is much longer than the duration of the entire process. Hence, this radiative decay is irrelevant. (The radiative decay into rovibrational levels of the same electronic state is irrelevant too, since these lifetimes are even longer \cite{Londono2023}.)

Finally, we have considered an isolated molecule. This approximation is well justified for sufficiently cold and low-density samples, where the time scales for vibrational relaxation can extend well beyond the nanosecond range \cite{Forrey1999}. Moreover, in this regime the process of population re-thermalization caused by the presence of black-body radiation takes a time of the order of seconds \cite{Leibfried2012}, so it can be neglected.

\subsection{Population dynamics}

\begin{figure}
    \centering
    \scalebox{0.355}{\includegraphics{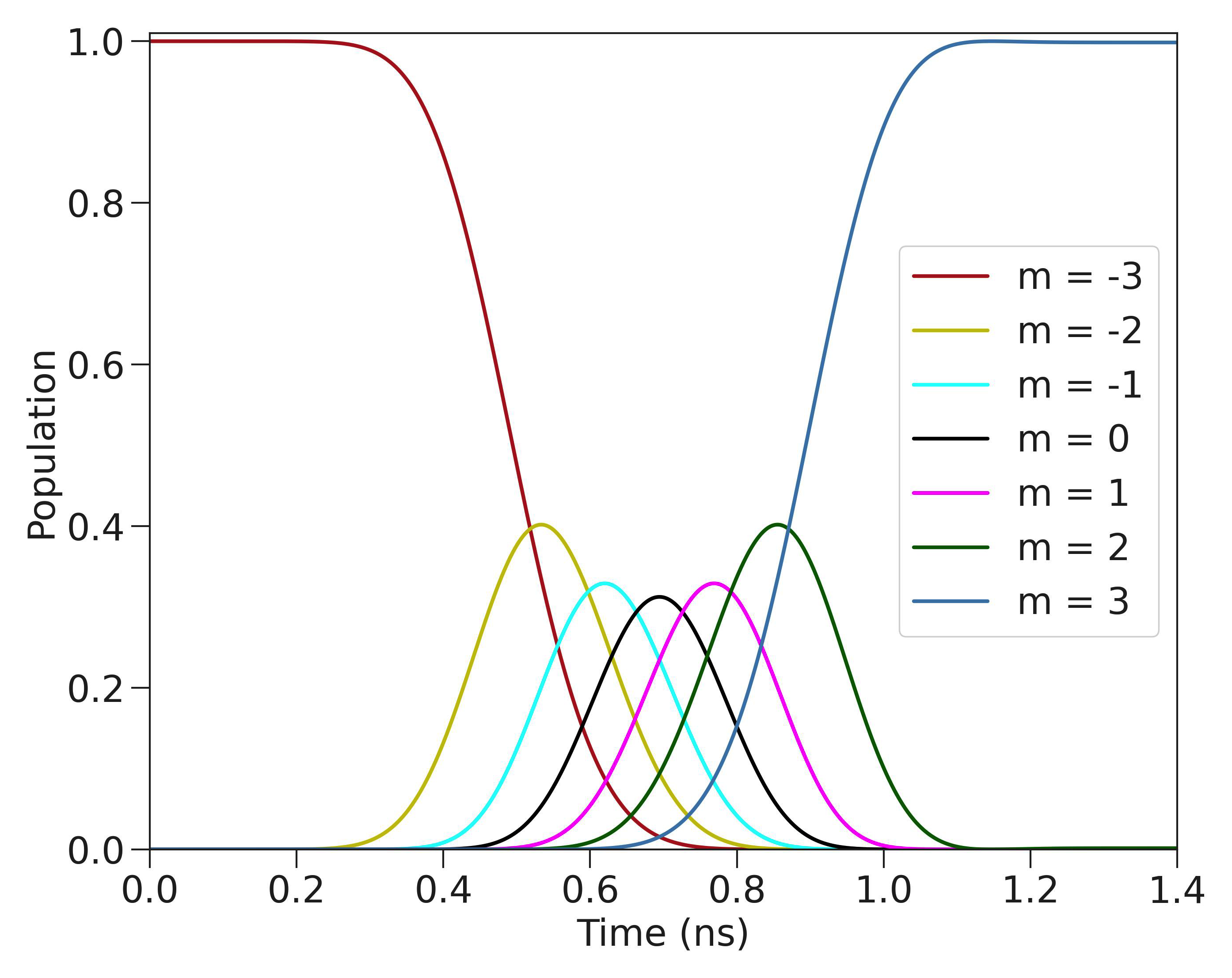}}
    \caption{Population dynamics of the states selected for the ping-pong rovibrational stabilization, obtained analytically by means of Eq. (\ref{eq:psm}).}
    \label{fig:model}
\end{figure}

Figure \ref{fig:model} displays the populations of the rovibronic states included in the chain, according to the analytical pseudospin model (\ref{eq:psm}) with the Rabi frequencies of Fig. \ref{fig:pulses}. It is observed that the initial state gets fully depopulated at about 0.8 ns and that at about 0.6 ns the target state begins to get population, until, at the end of the process, which occurs at about 1.1 ns, it becomes fully populated. Both behaviors are monotonic. Meanwhile, the intermediate states get consecutively populated and depopulated, with the maximum populations not surpassing 40\%. Clearly, the chain of states undergoes the ping-pong process, even though the laser pulses are applied simultaneously.

\begin{figure}
    \centering
    \scalebox{0.41}{\includegraphics{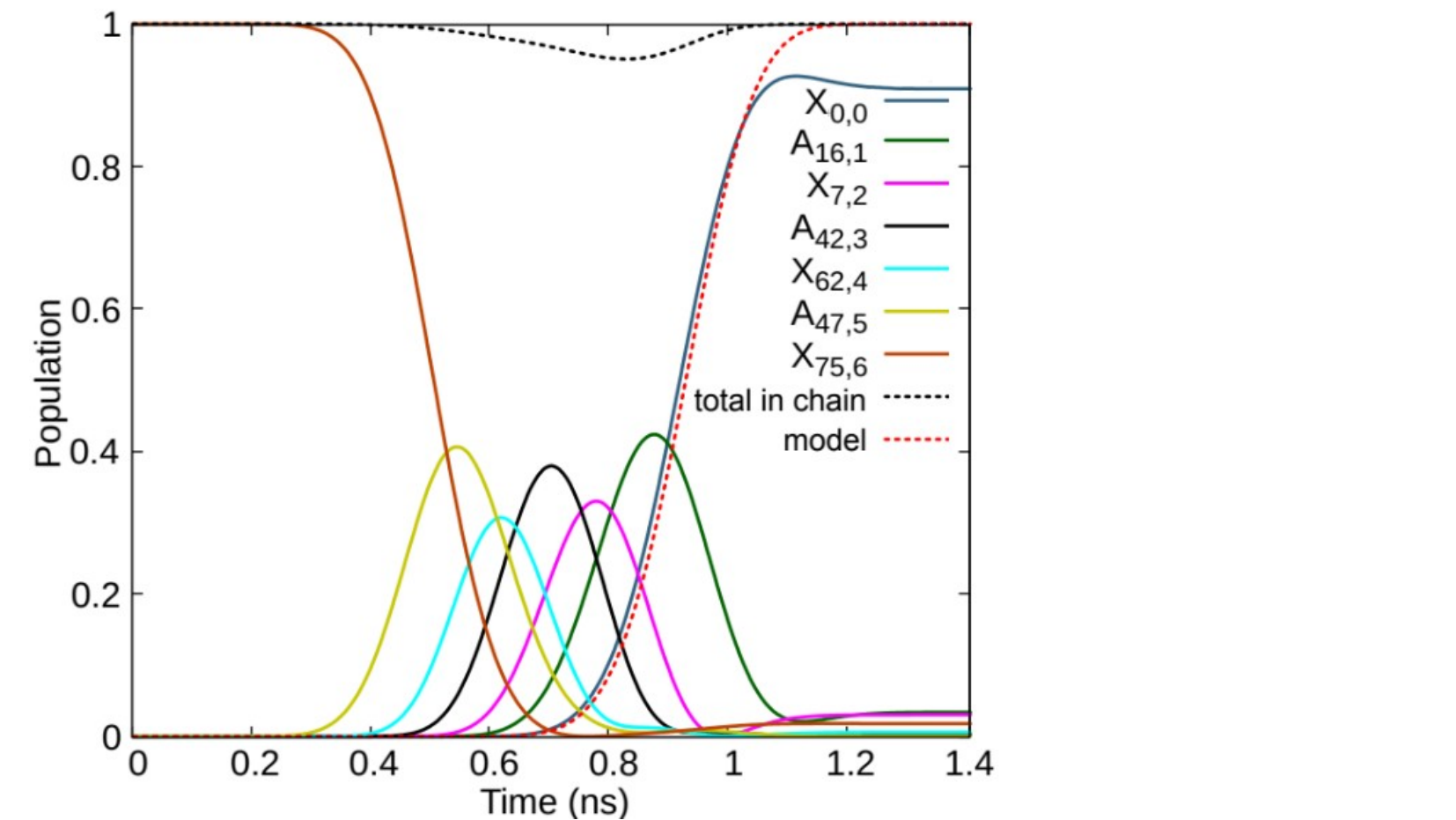}}
    \caption{Population dynamics of the states selected for the ping-pong rovibrational stabilization, obtained by numerical integration of Eqs. (\ref{eq:coupledeq}).}
    \label{fig:ocre_rpop}
\end{figure}

Figure \ref{fig:ocre_rpop} presents the populations obtained by means of the numerical integration of Eqs. (\ref{eq:coupledeq}). Evidently, the results are very similar to those of Fig. \ref{fig:model}, providing confidence in the approximations used in Sec. \ref{sec:RWA}. The main differences are that at about 1.1 ns the population of the target state exhibits a small shoulder, right before attaining a final value of 91\%, at the end of the process the initial and some of the intermediate states recover a little population, accounting for the remaining 9\%, and there occurs a small transient population leak out of the chain.

\section{Conclusions and Outlook}
\label{sec:conclusions}

The numerical solutions of the time-dependent Schr\"odinger equation show that the ping-pong scheme is sound. In addition, they demonstrate that the pseudospin, or SU($N$), is a faithful representation of the dynamics of this scheme.

This ping-pong scheme affords several advantages: First, it can be implemented with simple analytical laser pulses, available with current UV-Vis laser sources in the ps-ns range \cite{Oliveira2019, Zou2016}. Second, since it involves only rovibronic transitions, it can be applied to the rovibrational stabilization of nonpolar molecules, in contrast with other schemes that are confined to one potential energy curve \cite{Londono2023}. Third, it can deal with high densities of rovibrational levels. Finally, it can be applied to the preparation of an initially stable molecule in a selected excited rovibronic state, in particular, in a highly excited rovibrational level of the ground electronic state.

We plan to extend this scheme to include three or more potential energy curves. This will allow us, for example, to explicitly consider the spin-orbit coupling between excited electronic states and stabilize dimers formed in high-lying rovibrational levels of triplet electronic states \cite{Guerrero2018}.

\begin{acknowledgments}

This work was supported in part by Minciencias through Project No. 1106-658-42793. We thank CIBioFI of Universidad del Valle for kindly providing time in their computing facilities.

\end{acknowledgments}

\bibliography{Ref_article}

\begin{thebibliography}{47}%
\makeatletter
\providecommand \@ifxundefined [1]{%
 \@ifx{#1\undefined}
}%
\providecommand \@ifnum [1]{%
 \ifnum #1\expandafter \@firstoftwo
 \else \expandafter \@secondoftwo
 \fi
}%
\providecommand \@ifx [1]{%
 \ifx #1\expandafter \@firstoftwo
 \else \expandafter \@secondoftwo
 \fi
}%
\providecommand \natexlab [1]{#1}%
\providecommand \enquote  [1]{``#1''}%
\providecommand \bibnamefont  [1]{#1}%
\providecommand \bibfnamefont [1]{#1}%
\providecommand \citenamefont [1]{#1}%
\providecommand \href@noop [0]{\@secondoftwo}%
\providecommand \href [0]{\begingroup \@sanitize@url \@href}%
\providecommand \@href[1]{\@@startlink{#1}\@@href}%
\providecommand \@@href[1]{\endgroup#1\@@endlink}%
\providecommand \@sanitize@url [0]{\catcode `\\12\catcode `\$12\catcode `\&12\catcode `\#12\catcode `\^12\catcode `\_12\catcode `\%12\relax}%
\providecommand \@@startlink[1]{}%
\providecommand \@@endlink[0]{}%
\providecommand \url  [0]{\begingroup\@sanitize@url \@url }%
\providecommand \@url [1]{\endgroup\@href {#1}{\urlprefix }}%
\providecommand \urlprefix  [0]{URL }%
\providecommand \Eprint [0]{\href }%
\providecommand \doibase [0]{https://doi.org/}%
\providecommand \selectlanguage [0]{\@gobble}%
\providecommand \bibinfo  [0]{\@secondoftwo}%
\providecommand \bibfield  [0]{\@secondoftwo}%
\providecommand \translation [1]{[#1]}%
\providecommand \BibitemOpen [0]{}%
\providecommand \bibitemStop [0]{}%
\providecommand \bibitemNoStop [0]{.\EOS\space}%
\providecommand \EOS [0]{\spacefactor3000\relax}%
\providecommand \BibitemShut  [1]{\csname bibitem#1\endcsname}%
\let\auto@bib@innerbib\@empty
\bibitem [{\citenamefont {Weiner}\ \emph {et~al.}(1999)\citenamefont {Weiner}, \citenamefont {Bagnato}, \citenamefont {Zilio},\ and\ \citenamefont {Julienne}}]{Weiner1999}%
  \BibitemOpen
  \bibfield  {author} {\bibinfo {author} {\bibfnamefont {J.}~\bibnamefont {Weiner}}, \bibinfo {author} {\bibfnamefont {V.~S.}\ \bibnamefont {Bagnato}}, \bibinfo {author} {\bibfnamefont {S.}~\bibnamefont {Zilio}},\ and\ \bibinfo {author} {\bibfnamefont {P.~S.}\ \bibnamefont {Julienne}},\ }\href {https://doi.org/10.1103/RevModPhys.71.1} {\bibfield  {journal} {\bibinfo  {journal} {Rev. Mod. Phys.}\ }\textbf {\bibinfo {volume} {71}},\ \bibinfo {pages} {1} (\bibinfo {year} {1999})}\BibitemShut {NoStop}%
\bibitem [{\citenamefont {Carr}\ \emph {et~al.}(2009)\citenamefont {Carr}, \citenamefont {DeMille}, \citenamefont {Krems},\ and\ \citenamefont {Ye}}]{Carr2009}%
  \BibitemOpen
  \bibfield  {author} {\bibinfo {author} {\bibfnamefont {L.~D.}\ \bibnamefont {Carr}}, \bibinfo {author} {\bibfnamefont {D.}~\bibnamefont {DeMille}}, \bibinfo {author} {\bibfnamefont {R.~V.}\ \bibnamefont {Krems}},\ and\ \bibinfo {author} {\bibfnamefont {J.}~\bibnamefont {Ye}},\ }\href {https://doi.org/10.1088/1367-2630/11/5/055049} {\bibfield  {journal} {\bibinfo  {journal} {New J. Phys.}\ }\textbf {\bibinfo {volume} {11}},\ \bibinfo {pages} {055049} (\bibinfo {year} {2009})}\BibitemShut {NoStop}%
\bibitem [{\citenamefont {Bohn}\ \emph {et~al.}(2017)\citenamefont {Bohn}, \citenamefont {Rey},\ and\ \citenamefont {Ye}}]{Bohn2017}%
  \BibitemOpen
  \bibfield  {author} {\bibinfo {author} {\bibfnamefont {J.~L.}\ \bibnamefont {Bohn}}, \bibinfo {author} {\bibfnamefont {A.~M.}\ \bibnamefont {Rey}},\ and\ \bibinfo {author} {\bibfnamefont {J.}~\bibnamefont {Ye}},\ }\href {https://doi.org/10.1126/science.aam6299} {\bibfield  {journal} {\bibinfo  {journal} {Science}\ }\textbf {\bibinfo {volume} {357}},\ \bibinfo {pages} {1002} (\bibinfo {year} {2017})},\ \Eprint {https://arxiv.org/abs/1708.02806} {arXiv:1708.02806} \BibitemShut {NoStop}%
\bibitem [{\citenamefont {Dulieu}\ and\ \citenamefont {Osterwalder}(2018)}]{Dulieu2018}%
  \BibitemOpen
  \bibfield  {author} {\bibinfo {author} {\bibfnamefont {O.}~\bibnamefont {Dulieu}}\ and\ \bibinfo {author} {\bibfnamefont {A.}~\bibnamefont {Osterwalder}},\ }\href@noop {} {\emph {\bibinfo {title} {{Cold Chemistry: Molecular Scattering and Reactivity Near Absolute Zero}}}}\ (\bibinfo  {publisher} {The Royal Society of Chemistry},\ \bibinfo {address} {Croydon},\ \bibinfo {year} {2018})\BibitemShut {NoStop}%
\bibitem [{\citenamefont {R{\'\i}os}(2020)}]{perezriosbook}%
  \BibitemOpen
  \bibfield  {author} {\bibinfo {author} {\bibfnamefont {J.}~\bibnamefont {R{\'\i}os}},\ }in\ \href {https://books.google.com/books?id=pmcHEAAAQBAJ} {\emph {\bibinfo {booktitle} {An Introduction to Cold and Ultracold Chemistry: Atoms, Molecules, Ions and Rydbergs}}}\ (\bibinfo  {publisher} {Springer International Publishing},\ \bibinfo {year} {2020})\ Chap.\ \bibinfo {chapter} {Cold Chemical Reactions Between Molecular Ions and Neutral Atoms}, pp.\ \bibinfo {pages} {215 -- 234}\BibitemShut {NoStop}%
\bibitem [{\citenamefont {Blackmore}\ \emph {et~al.}(2018)\citenamefont {Blackmore}, \citenamefont {Caldwell}, \citenamefont {Gregory}, \citenamefont {Bridge}, \citenamefont {Sawant}, \citenamefont {Aldegunde}, \citenamefont {Mur-Petit}, \citenamefont {Jaksch}, \citenamefont {Hutson}, \citenamefont {Sauer}, \citenamefont {Tarbutt},\ and\ \citenamefont {Cornish}}]{Blackmore2019}%
  \BibitemOpen
  \bibfield  {author} {\bibinfo {author} {\bibfnamefont {J.~A.}\ \bibnamefont {Blackmore}}, \bibinfo {author} {\bibfnamefont {L.}~\bibnamefont {Caldwell}}, \bibinfo {author} {\bibfnamefont {P.~D.}\ \bibnamefont {Gregory}}, \bibinfo {author} {\bibfnamefont {E.~M.}\ \bibnamefont {Bridge}}, \bibinfo {author} {\bibfnamefont {R.}~\bibnamefont {Sawant}}, \bibinfo {author} {\bibfnamefont {J.}~\bibnamefont {Aldegunde}}, \bibinfo {author} {\bibfnamefont {J.}~\bibnamefont {Mur-Petit}}, \bibinfo {author} {\bibfnamefont {D.}~\bibnamefont {Jaksch}}, \bibinfo {author} {\bibfnamefont {J.~M.}\ \bibnamefont {Hutson}}, \bibinfo {author} {\bibfnamefont {B.~E.}\ \bibnamefont {Sauer}}, \bibinfo {author} {\bibfnamefont {M.~R.}\ \bibnamefont {Tarbutt}},\ and\ \bibinfo {author} {\bibfnamefont {S.~L.}\ \bibnamefont {Cornish}},\ }\href {https://doi.org/10.1088/2058-9565/aaee35} {\bibfield  {journal} {\bibinfo  {journal} {Quantum Sci. Technol.}\ }\textbf {\bibinfo {volume} {4}},\ \bibinfo {pages} {014010} (\bibinfo {year}
  {2018})}\BibitemShut {NoStop}%
\bibitem [{\citenamefont {Schuster}\ \emph {et~al.}(2021)\citenamefont {Schuster}, \citenamefont {Flicker}, \citenamefont {Li}, \citenamefont {Kotochigova}, \citenamefont {Moore}, \citenamefont {Ye},\ and\ \citenamefont {Yao}}]{Schuster2021}%
  \BibitemOpen
  \bibfield  {author} {\bibinfo {author} {\bibfnamefont {T.}~\bibnamefont {Schuster}}, \bibinfo {author} {\bibfnamefont {F.}~\bibnamefont {Flicker}}, \bibinfo {author} {\bibfnamefont {M.}~\bibnamefont {Li}}, \bibinfo {author} {\bibfnamefont {S.}~\bibnamefont {Kotochigova}}, \bibinfo {author} {\bibfnamefont {J.~E.}\ \bibnamefont {Moore}}, \bibinfo {author} {\bibfnamefont {J.}~\bibnamefont {Ye}},\ and\ \bibinfo {author} {\bibfnamefont {N.~Y.}\ \bibnamefont {Yao}},\ }\href {https://doi.org/10.1103/PhysRevA.103.063322} {\bibfield  {journal} {\bibinfo  {journal} {Phys. Rev. A}\ }\textbf {\bibinfo {volume} {103}},\ \bibinfo {pages} {063322} (\bibinfo {year} {2021})}\BibitemShut {NoStop}%
\bibitem [{\citenamefont {Hughes}\ \emph {et~al.}(2023)\citenamefont {Hughes}, \citenamefont {Lode}, \citenamefont {Jaksch},\ and\ \citenamefont {Molignini}}]{Hughes2023}%
  \BibitemOpen
  \bibfield  {author} {\bibinfo {author} {\bibfnamefont {M.}~\bibnamefont {Hughes}}, \bibinfo {author} {\bibfnamefont {A.~U.~J.}\ \bibnamefont {Lode}}, \bibinfo {author} {\bibfnamefont {D.}~\bibnamefont {Jaksch}},\ and\ \bibinfo {author} {\bibfnamefont {P.}~\bibnamefont {Molignini}},\ }\href {https://doi.org/10.1103/PhysRevA.107.033323} {\bibfield  {journal} {\bibinfo  {journal} {Phys. Rev. A}\ }\textbf {\bibinfo {volume} {107}},\ \bibinfo {pages} {033323} (\bibinfo {year} {2023})}\BibitemShut {NoStop}%
\bibitem [{\citenamefont {Liu}\ \emph {et~al.}(2021)\citenamefont {Liu}, \citenamefont {Hu}, \citenamefont {Nichols}, \citenamefont {Yang}, \citenamefont {Xie}, \citenamefont {Guo},\ and\ \citenamefont {Ni}}]{Ni2021}%
  \BibitemOpen
  \bibfield  {author} {\bibinfo {author} {\bibfnamefont {Y.}~\bibnamefont {Liu}}, \bibinfo {author} {\bibfnamefont {M.-G.}\ \bibnamefont {Hu}}, \bibinfo {author} {\bibfnamefont {M.~A.}\ \bibnamefont {Nichols}}, \bibinfo {author} {\bibfnamefont {D.}~\bibnamefont {Yang}}, \bibinfo {author} {\bibfnamefont {D.}~\bibnamefont {Xie}}, \bibinfo {author} {\bibfnamefont {H.}~\bibnamefont {Guo}},\ and\ \bibinfo {author} {\bibfnamefont {K.-K.}\ \bibnamefont {Ni}},\ }\href {https://www.osti.gov/biblio/1830683} {\bibfield  {journal} {\bibinfo  {journal} {Nature (London)}\ }\textbf {\bibinfo {volume} {593}} (\bibinfo {year} {2021})}\BibitemShut {NoStop}%
\bibitem [{\citenamefont {Liu}\ and\ \citenamefont {Ni}(2022)}]{Ni2022}%
  \BibitemOpen
  \bibfield  {author} {\bibinfo {author} {\bibfnamefont {Y.}~\bibnamefont {Liu}}\ and\ \bibinfo {author} {\bibfnamefont {K.-K.}\ \bibnamefont {Ni}},\ }\href {https://doi.org/10.1146/annurev-physchem-090419-043244} {\bibfield  {journal} {\bibinfo  {journal} {Annu. Rev. Phys. Chem.}\ }\textbf {\bibinfo {volume} {73}},\ \bibinfo {pages} {73} (\bibinfo {year} {2022})}\BibitemShut {NoStop}%
\bibitem [{\citenamefont {Deng}\ \emph {et~al.}(2023)\citenamefont {Deng}, \citenamefont {Chen}, \citenamefont {Luo}, \citenamefont {Zhang}, \citenamefont {Yi},\ and\ \citenamefont {Shi}}]{Deng2023}%
  \BibitemOpen
  \bibfield  {author} {\bibinfo {author} {\bibfnamefont {F.}~\bibnamefont {Deng}}, \bibinfo {author} {\bibfnamefont {X.-Y.}\ \bibnamefont {Chen}}, \bibinfo {author} {\bibfnamefont {X.-Y.}\ \bibnamefont {Luo}}, \bibinfo {author} {\bibfnamefont {W.}~\bibnamefont {Zhang}}, \bibinfo {author} {\bibfnamefont {S.}~\bibnamefont {Yi}},\ and\ \bibinfo {author} {\bibfnamefont {T.}~\bibnamefont {Shi}},\ }\href {https://doi.org/10.1103/PhysRevLett.130.183001} {\bibfield  {journal} {\bibinfo  {journal} {Phys. Rev. Lett.}\ }\textbf {\bibinfo {volume} {130}},\ \bibinfo {pages} {183001} (\bibinfo {year} {2023})}\BibitemShut {NoStop}%
\bibitem [{\citenamefont {Duda}\ \emph {et~al.}(2023)\citenamefont {Duda}, \citenamefont {Chen}, \citenamefont {Schindewolf}, \citenamefont {Bause}, \citenamefont {Milczewski}, \citenamefont {Schmidt}, \citenamefont {Bloch},\ and\ \citenamefont {Luo}}]{Duda2023}%
  \BibitemOpen
  \bibfield  {author} {\bibinfo {author} {\bibfnamefont {M.}~\bibnamefont {Duda}}, \bibinfo {author} {\bibfnamefont {X.-Y.}\ \bibnamefont {Chen}}, \bibinfo {author} {\bibfnamefont {A.}~\bibnamefont {Schindewolf}}, \bibinfo {author} {\bibfnamefont {R.}~\bibnamefont {Bause}}, \bibinfo {author} {\bibfnamefont {J.}~\bibnamefont {Milczewski}}, \bibinfo {author} {\bibfnamefont {R.}~\bibnamefont {Schmidt}}, \bibinfo {author} {\bibfnamefont {I.}~\bibnamefont {Bloch}},\ and\ \bibinfo {author} {\bibfnamefont {X.}~\bibnamefont {Luo}},\ }\href {https://doi.org/10.1038/s41567-023-01948-1} {\bibfield  {journal} {\bibinfo  {journal} {Nat. Phys.}\ }\textbf {\bibinfo {volume} {19}},\ \bibinfo {pages} {1} (\bibinfo {year} {2023})}\BibitemShut {NoStop}%
\bibitem [{\citenamefont {Parazzoli}\ \emph {et~al.}(2009)\citenamefont {Parazzoli}, \citenamefont {Fitch}, \citenamefont {Lobser},\ and\ \citenamefont {Lewandowski}}]{Parazzoli2009}%
  \BibitemOpen
  \bibfield  {author} {\bibinfo {author} {\bibfnamefont {L.~P.}\ \bibnamefont {Parazzoli}}, \bibinfo {author} {\bibfnamefont {N.}~\bibnamefont {Fitch}}, \bibinfo {author} {\bibfnamefont {D.~S.}\ \bibnamefont {Lobser}},\ and\ \bibinfo {author} {\bibfnamefont {H.~J.}\ \bibnamefont {Lewandowski}},\ }\href {https://doi.org/10.1088/1367-2630/11/5/055031} {\bibfield  {journal} {\bibinfo  {journal} {New J. Phys.}\ }\textbf {\bibinfo {volume} {11}},\ \bibinfo {pages} {055031} (\bibinfo {year} {2009})}\BibitemShut {NoStop}%
\bibitem [{\citenamefont {Krems.}\ \emph {et~al.}(2009)\citenamefont {Krems.}, \citenamefont {Stwalley},\ and\ \citenamefont {Friedrich}}]{Krems.2009}%
  \BibitemOpen
  \bibinfo {editor} {\bibfnamefont {R.~V.}\ \bibnamefont {Krems.}}, \bibinfo {editor} {\bibfnamefont {W.~C.}\ \bibnamefont {Stwalley}},\ and\ \bibinfo {editor} {\bibfnamefont {B.}~\bibnamefont {Friedrich}},\ eds.,\ \href@noop {} {\emph {\bibinfo {title} {{Cold Molecules, Theory, Experiements, Applications}}}}\ (\bibinfo  {publisher} {CRC Press Taylor and Francis Group},\ \bibinfo {address} {New York},\ \bibinfo {year} {2009})\ p.\ \bibinfo {pages} {751}\BibitemShut {NoStop}%
\bibitem [{\citenamefont {Manai}\ \emph {et~al.}(2012)\citenamefont {Manai}, \citenamefont {Horchani}, \citenamefont {Lignier}, \citenamefont {Pillet}, \citenamefont {Comparat}, \citenamefont {Fioretti},\ and\ \citenamefont {Allegrini}}]{Manai2012}%
  \BibitemOpen
  \bibfield  {author} {\bibinfo {author} {\bibfnamefont {I.}~\bibnamefont {Manai}}, \bibinfo {author} {\bibfnamefont {R.}~\bibnamefont {Horchani}}, \bibinfo {author} {\bibfnamefont {H.}~\bibnamefont {Lignier}}, \bibinfo {author} {\bibfnamefont {P.}~\bibnamefont {Pillet}}, \bibinfo {author} {\bibfnamefont {D.}~\bibnamefont {Comparat}}, \bibinfo {author} {\bibfnamefont {A.}~\bibnamefont {Fioretti}},\ and\ \bibinfo {author} {\bibfnamefont {M.}~\bibnamefont {Allegrini}},\ }\href {https://doi.org/10.1103/PhysRevLett.109.183001} {\bibfield  {journal} {\bibinfo  {journal} {Phys. Rev. Lett.}\ }\textbf {\bibinfo {volume} {109}},\ \bibinfo {pages} {183001} (\bibinfo {year} {2012})}\BibitemShut {NoStop}%
\bibitem [{\citenamefont {Bigelow}(2012)}]{Bigelow2012}%
  \BibitemOpen
  \bibfield  {author} {\bibinfo {author} {\bibfnamefont {N.}~\bibnamefont {Bigelow}},\ }\href {https://doi.org/10.1103/physics.5.121} {\bibfield  {journal} {\bibinfo  {journal} {Physics}\ }\textbf {\bibinfo {volume} {5}},\ \bibinfo {pages} {121} (\bibinfo {year} {2012})}\BibitemShut {NoStop}%
\bibitem [{\citenamefont {Lemeshko}\ \emph {et~al.}(2013)\citenamefont {Lemeshko}, \citenamefont {Krems}, \citenamefont {Doyle},\ and\ \citenamefont {Kais}}]{Lemeshko2013}%
  \BibitemOpen
  \bibfield  {author} {\bibinfo {author} {\bibfnamefont {M.}~\bibnamefont {Lemeshko}}, \bibinfo {author} {\bibfnamefont {R.~V.}\ \bibnamefont {Krems}}, \bibinfo {author} {\bibfnamefont {J.~M.}\ \bibnamefont {Doyle}},\ and\ \bibinfo {author} {\bibfnamefont {S.}~\bibnamefont {Kais}},\ }\href {https://doi.org/10.1080/00268976.2013.813595} {\bibfield  {journal} {\bibinfo  {journal} {Mol. Phys.}\ }\textbf {\bibinfo {volume} {111}},\ \bibinfo {pages} {1648} (\bibinfo {year} {2013})}\BibitemShut {NoStop}%
\bibitem [{\citenamefont {Hamamda}\ \emph {et~al.}(2015)\citenamefont {Hamamda}, \citenamefont {Pillet}, \citenamefont {Lignier},\ and\ \citenamefont {Comparat}}]{Hamamda2015}%
  \BibitemOpen
  \bibfield  {author} {\bibinfo {author} {\bibfnamefont {M.}~\bibnamefont {Hamamda}}, \bibinfo {author} {\bibfnamefont {P.}~\bibnamefont {Pillet}}, \bibinfo {author} {\bibfnamefont {H.}~\bibnamefont {Lignier}},\ and\ \bibinfo {author} {\bibfnamefont {D.}~\bibnamefont {Comparat}},\ }\href {https://doi.org/10.1088/0953-4075/48/18/182001} {\bibfield  {journal} {\bibinfo  {journal} {J. Phys. B: At. Mol. Opt. Phys.}\ }\textbf {\bibinfo {volume} {48}},\ \bibinfo {pages} {22796} (\bibinfo {year} {2015})}\BibitemShut {NoStop}%
\bibitem [{\citenamefont {Marcelis}\ \emph {et~al.}(2008)\citenamefont {Marcelis}, \citenamefont {Verhaar},\ and\ \citenamefont {Kokkelmans}}]{Marcelis2008}%
  \BibitemOpen
  \bibfield  {author} {\bibinfo {author} {\bibfnamefont {B.}~\bibnamefont {Marcelis}}, \bibinfo {author} {\bibfnamefont {B.}~\bibnamefont {Verhaar}},\ and\ \bibinfo {author} {\bibfnamefont {S.}~\bibnamefont {Kokkelmans}},\ }\href {https://doi.org/10.1103/PhysRevLett.100.153201} {\bibfield  {journal} {\bibinfo  {journal} {Phys. Rev. Lett.}\ }\textbf {\bibinfo {volume} {100}},\ \bibinfo {pages} {153201} (\bibinfo {year} {2008})}\BibitemShut {NoStop}%
\bibitem [{\citenamefont {Thalhammer}\ \emph {et~al.}(2009)\citenamefont {Thalhammer}, \citenamefont {Barontini}, \citenamefont {Catani}, \citenamefont {Rabatti}, \citenamefont {Weber}, \citenamefont {Simoni}, \citenamefont {Minardi},\ and\ \citenamefont {Lnguscio}}]{Thalhammer2009}%
  \BibitemOpen
  \bibfield  {author} {\bibinfo {author} {\bibfnamefont {G.}~\bibnamefont {Thalhammer}}, \bibinfo {author} {\bibfnamefont {G.}~\bibnamefont {Barontini}}, \bibinfo {author} {\bibfnamefont {J.}~\bibnamefont {Catani}}, \bibinfo {author} {\bibfnamefont {F.}~\bibnamefont {Rabatti}}, \bibinfo {author} {\bibfnamefont {C.}~\bibnamefont {Weber}}, \bibinfo {author} {\bibfnamefont {A.}~\bibnamefont {Simoni}}, \bibinfo {author} {\bibfnamefont {F.}~\bibnamefont {Minardi}},\ and\ \bibinfo {author} {\bibfnamefont {M.}~\bibnamefont {Lnguscio}},\ }\href {https://doi.org/10.1088/1367-2630/11/5/055044} {\bibfield  {journal} {\bibinfo  {journal} {New J. Phys.}\ }\textbf {\bibinfo {volume} {11}},\ \bibinfo {pages} {055044} (\bibinfo {year} {2009})}\BibitemShut {NoStop}%
\bibitem [{\citenamefont {Chin}\ \emph {et~al.}(2010)\citenamefont {Chin}, \citenamefont {Grimm}, \citenamefont {Julienne},\ and\ \citenamefont {Tiesinga}}]{Chin2010}%
  \BibitemOpen
  \bibfield  {author} {\bibinfo {author} {\bibfnamefont {C.}~\bibnamefont {Chin}}, \bibinfo {author} {\bibfnamefont {R.}~\bibnamefont {Grimm}}, \bibinfo {author} {\bibfnamefont {P.}~\bibnamefont {Julienne}},\ and\ \bibinfo {author} {\bibfnamefont {E.}~\bibnamefont {Tiesinga}},\ }\href {https://doi.org/10.1103/RevModPhys.82.1225} {\bibfield  {journal} {\bibinfo  {journal} {Rev. Mod. Phys.}\ }\textbf {\bibinfo {volume} {82}},\ \bibinfo {pages} {1225} (\bibinfo {year} {2010})}\BibitemShut {NoStop}%
\bibitem [{\citenamefont {Wacker}\ \emph {et~al.}(2015)\citenamefont {Wacker}, \citenamefont {J{\o}rgensen}, \citenamefont {Birkmose}, \citenamefont {Horchani}, \citenamefont {Ertmer}, \citenamefont {Klempt}, \citenamefont {Winter}, \citenamefont {Sherson},\ and\ \citenamefont {Arlt}}]{Wacker2015}%
  \BibitemOpen
  \bibfield  {author} {\bibinfo {author} {\bibfnamefont {L.}~\bibnamefont {Wacker}}, \bibinfo {author} {\bibfnamefont {N.~B.}\ \bibnamefont {J{\o}rgensen}}, \bibinfo {author} {\bibfnamefont {D.}~\bibnamefont {Birkmose}}, \bibinfo {author} {\bibfnamefont {R.}~\bibnamefont {Horchani}}, \bibinfo {author} {\bibfnamefont {W.}~\bibnamefont {Ertmer}}, \bibinfo {author} {\bibfnamefont {C.}~\bibnamefont {Klempt}}, \bibinfo {author} {\bibfnamefont {N.}~\bibnamefont {Winter}}, \bibinfo {author} {\bibfnamefont {J.}~\bibnamefont {Sherson}},\ and\ \bibinfo {author} {\bibfnamefont {J.~J.}\ \bibnamefont {Arlt}},\ }\href {https://doi.org/10.1103/PhysRevA.92.053602} {\bibfield  {journal} {\bibinfo  {journal} {Phys. Rev. A}\ }\textbf {\bibinfo {volume} {92}},\ \bibinfo {pages} {053602} (\bibinfo {year} {2015})}\BibitemShut {NoStop}%
\bibitem [{\citenamefont {Ulmanis}\ \emph {et~al.}(2012)\citenamefont {Ulmanis}, \citenamefont {Deiglmayr}, \citenamefont {Repp}, \citenamefont {Wester},\ and\ \citenamefont {Weidem{\"{u}}ller}}]{Ulmanis2012}%
  \BibitemOpen
  \bibfield  {author} {\bibinfo {author} {\bibfnamefont {J.}~\bibnamefont {Ulmanis}}, \bibinfo {author} {\bibfnamefont {J.}~\bibnamefont {Deiglmayr}}, \bibinfo {author} {\bibfnamefont {M.}~\bibnamefont {Repp}}, \bibinfo {author} {\bibfnamefont {R.}~\bibnamefont {Wester}},\ and\ \bibinfo {author} {\bibfnamefont {M.}~\bibnamefont {Weidem{\"{u}}ller}},\ }\href {https://doi.org/10.1021/cr300215h} {\bibfield  {journal} {\bibinfo  {journal} {Chem. Rev.}\ }\textbf {\bibinfo {volume} {112}},\ \bibinfo {pages} {4890} (\bibinfo {year} {2012})}\BibitemShut {NoStop}%
\bibitem [{\citenamefont {Molano}\ \emph {et~al.}(2019)\citenamefont {Molano}, \citenamefont {P{\'{e}}rez}, \citenamefont {Arce}, \citenamefont {L{\'{o}}pez},\ and\ \citenamefont {Zambrano}}]{Molano2019}%
  \BibitemOpen
  \bibfield  {author} {\bibinfo {author} {\bibfnamefont {J.~S.}\ \bibnamefont {Molano}}, \bibinfo {author} {\bibfnamefont {K.~D.}\ \bibnamefont {P{\'{e}}rez}}, \bibinfo {author} {\bibfnamefont {J.~C.}\ \bibnamefont {Arce}}, \bibinfo {author} {\bibfnamefont {J.~G.}\ \bibnamefont {L{\'{o}}pez}},\ and\ \bibinfo {author} {\bibfnamefont {M.~L.}\ \bibnamefont {Zambrano}},\ }\href {https://doi.org/10.1103/PhysRevA.100.063407} {\bibfield  {journal} {\bibinfo  {journal} {Phys. Rev. A}\ }\textbf {\bibinfo {volume} {100}},\ \bibinfo {pages} {063407} (\bibinfo {year} {2019})}\BibitemShut {NoStop}%
\bibitem [{\citenamefont {Eisele}\ \emph {et~al.}(2020)\citenamefont {Eisele}, \citenamefont {Maier},\ and\ \citenamefont {Zimmermann}}]{Eisele2020}%
  \BibitemOpen
  \bibfield  {author} {\bibinfo {author} {\bibfnamefont {M.}~\bibnamefont {Eisele}}, \bibinfo {author} {\bibfnamefont {R.~A.}\ \bibnamefont {Maier}},\ and\ \bibinfo {author} {\bibfnamefont {C.}~\bibnamefont {Zimmermann}},\ }\href {https://doi.org/10.1103/PhysRevLett.124.123401} {\bibfield  {journal} {\bibinfo  {journal} {Phys. Rev. Lett.}\ }\textbf {\bibinfo {volume} {124}},\ \bibinfo {pages} {123401} (\bibinfo {year} {2020})}\BibitemShut {NoStop}%
\bibitem [{\citenamefont {Stevenson}\ \emph {et~al.}(2016)\citenamefont {Stevenson}, \citenamefont {Blasing}, \citenamefont {Chen},\ and\ \citenamefont {Elliott}}]{Stevenson2016}%
  \BibitemOpen
  \bibfield  {author} {\bibinfo {author} {\bibfnamefont {I.~C.}\ \bibnamefont {Stevenson}}, \bibinfo {author} {\bibfnamefont {D.~B.}\ \bibnamefont {Blasing}}, \bibinfo {author} {\bibfnamefont {Y.~P.}\ \bibnamefont {Chen}},\ and\ \bibinfo {author} {\bibfnamefont {D.~S.}\ \bibnamefont {Elliott}},\ }\href {https://doi.org/10.1103/PhysRevA.94.062510} {\bibfield  {journal} {\bibinfo  {journal} {Phys. Rev. A}\ }\textbf {\bibinfo {volume} {94}},\ \bibinfo {pages} {062510} (\bibinfo {year} {2016})}\BibitemShut {NoStop}%
\bibitem [{\citenamefont {Aikawa}\ \emph {et~al.}(2010)\citenamefont {Aikawa}, \citenamefont {Akamatsu}, \citenamefont {Hayashi}, \citenamefont {Oasa}, \citenamefont {Kobayashi}, \citenamefont {Naidon}, \citenamefont {Kishimoto}, \citenamefont {Ueda},\ and\ \citenamefont {Inouye}}]{Aikawa2010}%
  \BibitemOpen
  \bibfield  {author} {\bibinfo {author} {\bibfnamefont {K.}~\bibnamefont {Aikawa}}, \bibinfo {author} {\bibfnamefont {D.}~\bibnamefont {Akamatsu}}, \bibinfo {author} {\bibfnamefont {M.}~\bibnamefont {Hayashi}}, \bibinfo {author} {\bibfnamefont {K.}~\bibnamefont {Oasa}}, \bibinfo {author} {\bibfnamefont {J.}~\bibnamefont {Kobayashi}}, \bibinfo {author} {\bibfnamefont {P.}~\bibnamefont {Naidon}}, \bibinfo {author} {\bibfnamefont {T.}~\bibnamefont {Kishimoto}}, \bibinfo {author} {\bibfnamefont {M.}~\bibnamefont {Ueda}},\ and\ \bibinfo {author} {\bibfnamefont {S.}~\bibnamefont {Inouye}},\ }\href {https://doi.org/10.1103/PhysRevLett.105.203001} {\bibfield  {journal} {\bibinfo  {journal} {Phys. Rev. Lett.}\ }\textbf {\bibinfo {volume} {105}},\ \bibinfo {pages} {203001} (\bibinfo {year} {2010})}\BibitemShut {NoStop}%
\bibitem [{\citenamefont {Koch}\ and\ \citenamefont {Shapiro}(2012)}]{Koch2012}%
  \BibitemOpen
  \bibfield  {author} {\bibinfo {author} {\bibfnamefont {C.~P.}\ \bibnamefont {Koch}}\ and\ \bibinfo {author} {\bibfnamefont {M.}~\bibnamefont {Shapiro}},\ }\href {https://doi.org/10.1021/cr2003882} {\bibfield  {journal} {\bibinfo  {journal} {Chem. Rev.}\ }\textbf {\bibinfo {volume} {112}},\ \bibinfo {pages} {4928} (\bibinfo {year} {2012})}\BibitemShut {NoStop}%
\bibitem [{\citenamefont {Salzmann}(2007)}]{Salzmann2007}%
  \BibitemOpen
  \bibfield  {author} {\bibinfo {author} {\bibfnamefont {W.}~\bibnamefont {Salzmann}},\ }\emph {\bibinfo {title} {{Photoassociation and coherent control of ultracold molecules by femtosecond pulses}}},\ \href@noop {} {Ph.D. thesis},\ \bibinfo  {school} {Albert-Ludwigs-universität} (\bibinfo {year} {2007})\BibitemShut {NoStop}%
\bibitem [{\citenamefont {Wang}\ \emph {et~al.}(2017)\citenamefont {Wang}, \citenamefont {Li}, \citenamefont {Hu}, \citenamefont {Chen},\ and\ \citenamefont {Cong}}]{Wang2017}%
  \BibitemOpen
  \bibfield  {author} {\bibinfo {author} {\bibfnamefont {M.}~\bibnamefont {Wang}}, \bibinfo {author} {\bibfnamefont {J.~L.}\ \bibnamefont {Li}}, \bibinfo {author} {\bibfnamefont {X.~J.}\ \bibnamefont {Hu}}, \bibinfo {author} {\bibfnamefont {M.~D.}\ \bibnamefont {Chen}},\ and\ \bibinfo {author} {\bibfnamefont {S.~L.}\ \bibnamefont {Cong}},\ }\href {https://doi.org/10.1103/PhysRevA.96.043417} {\bibfield  {journal} {\bibinfo  {journal} {Phys. Rev. A}\ }\textbf {\bibinfo {volume} {96}},\ \bibinfo {pages} {043417} (\bibinfo {year} {2017})}\BibitemShut {NoStop}%
\bibitem [{\citenamefont {Carini}\ \emph {et~al.}(2016)\citenamefont {Carini}, \citenamefont {Kallush}, \citenamefont {Kosloff},\ and\ \citenamefont {Gould}}]{Carini2016}%
  \BibitemOpen
  \bibfield  {author} {\bibinfo {author} {\bibfnamefont {J.~L.}\ \bibnamefont {Carini}}, \bibinfo {author} {\bibfnamefont {S.}~\bibnamefont {Kallush}}, \bibinfo {author} {\bibfnamefont {R.}~\bibnamefont {Kosloff}},\ and\ \bibinfo {author} {\bibfnamefont {P.~L.}\ \bibnamefont {Gould}},\ }\href {https://doi.org/10.1021/acs.jpca.5b10088} {\bibfield  {journal} {\bibinfo  {journal} {J. Phys. Chem. A}\ }\textbf {\bibinfo {volume} {120}},\ \bibinfo {pages} {3032} (\bibinfo {year} {2016})}\BibitemShut {NoStop}%
\bibitem [{\citenamefont {Ni}\ \emph {et~al.}(2008)\citenamefont {Ni}, \citenamefont {Ospelkaus}, \citenamefont {{De Miranda}}, \citenamefont {Pe'er}, \citenamefont {Neyenhuis}, \citenamefont {Zirbel}, \citenamefont {Kotochigova}, \citenamefont {Julienne}, \citenamefont {Jin},\ and\ \citenamefont {Ye}}]{Ni2008}%
  \BibitemOpen
  \bibfield  {author} {\bibinfo {author} {\bibfnamefont {K.~K.}\ \bibnamefont {Ni}}, \bibinfo {author} {\bibfnamefont {S.}~\bibnamefont {Ospelkaus}}, \bibinfo {author} {\bibfnamefont {M.~H.}\ \bibnamefont {{De Miranda}}}, \bibinfo {author} {\bibfnamefont {A.}~\bibnamefont {Pe'er}}, \bibinfo {author} {\bibfnamefont {B.}~\bibnamefont {Neyenhuis}}, \bibinfo {author} {\bibfnamefont {J.~J.}\ \bibnamefont {Zirbel}}, \bibinfo {author} {\bibfnamefont {S.}~\bibnamefont {Kotochigova}}, \bibinfo {author} {\bibfnamefont {P.~S.}\ \bibnamefont {Julienne}}, \bibinfo {author} {\bibfnamefont {D.~S.}\ \bibnamefont {Jin}},\ and\ \bibinfo {author} {\bibfnamefont {J.}~\bibnamefont {Ye}},\ }\href {https://doi.org/10.1126/science.1163861} {\bibfield  {journal} {\bibinfo  {journal} {Science}\ }\textbf {\bibinfo {volume} {322}},\ \bibinfo {pages} {231} (\bibinfo {year} {2008})}\BibitemShut {NoStop}%
\bibitem [{\citenamefont {{De Lima}}(2017)}]{DeLima2017}%
  \BibitemOpen
  \bibfield  {author} {\bibinfo {author} {\bibfnamefont {E.~F.}\ \bibnamefont {{De Lima}}},\ }\href {https://doi.org/10.1103/PhysRevA.95.013411} {\bibfield  {journal} {\bibinfo  {journal} {Phys. Rev. A}\ }\textbf {\bibinfo {volume} {95}},\ \bibinfo {pages} {013411} (\bibinfo {year} {2017})}\BibitemShut {NoStop}%
\bibitem [{\citenamefont {de~Lima}(2015)}]{DeLima2015}%
  \BibitemOpen
  \bibfield  {author} {\bibinfo {author} {\bibfnamefont {E.~F.}\ \bibnamefont {de~Lima}},\ }\href {https://doi.org/10.1007/s10909-015-1280-3} {\bibfield  {journal} {\bibinfo  {journal} {J. Low Temp. Phys.}\ }\textbf {\bibinfo {volume} {180}},\ \bibinfo {pages} {161} (\bibinfo {year} {2015})}\BibitemShut {NoStop}%
\bibitem [{\citenamefont {Ndong}\ and\ \citenamefont {Koch}(2010)}]{Ndong2010}%
  \BibitemOpen
  \bibfield  {author} {\bibinfo {author} {\bibfnamefont {M.}~\bibnamefont {Ndong}}\ and\ \bibinfo {author} {\bibfnamefont {C.~P.}\ \bibnamefont {Koch}},\ }\href {https://doi.org/10.1103/PhysRevA.82.043437} {\bibfield  {journal} {\bibinfo  {journal} {Phys. Rev. A}\ }\textbf {\bibinfo {volume} {82}},\ \bibinfo {pages} {1} (\bibinfo {year} {2010})}\BibitemShut {NoStop}%
\bibitem [{\citenamefont {Guerrero}\ \emph {et~al.}(2018)\citenamefont {Guerrero}, \citenamefont {Castellanos},\ and\ \citenamefont {Arango}}]{Guerrero2018}%
  \BibitemOpen
  \bibfield  {author} {\bibinfo {author} {\bibfnamefont {R.~D.}\ \bibnamefont {Guerrero}}, \bibinfo {author} {\bibfnamefont {M.~A.}\ \bibnamefont {Castellanos}},\ and\ \bibinfo {author} {\bibfnamefont {C.~A.}\ \bibnamefont {Arango}},\ }\bibfield  {journal} {\bibinfo  {journal} {J. Chem. Phys.}\ }\textbf {\bibinfo {volume} {149}},\ \href {https://doi.org/10.1063/1.5052019} {10.1063/1.5052019} (\bibinfo {year} {2018})\BibitemShut {NoStop}%
\bibitem [{\citenamefont {Londo\~no}\ and\ \citenamefont {Arce}(2023)}]{Londono2023}%
  \BibitemOpen
  \bibfield  {author} {\bibinfo {author} {\bibfnamefont {M.}~\bibnamefont {Londo\~no}}\ and\ \bibinfo {author} {\bibfnamefont {J.~C.}\ \bibnamefont {Arce}},\ }\href {https://doi.org/10.1103/PhysRevA.108.013301} {\bibfield  {journal} {\bibinfo  {journal} {Phys. Rev. A}\ }\textbf {\bibinfo {volume} {108}},\ \bibinfo {pages} {013301} (\bibinfo {year} {2023})}\BibitemShut {NoStop}%
\bibitem [{\citenamefont {Gonz{\'{a}}lez-F{\'{e}}rez}\ and\ \citenamefont {Koch}(2012)}]{Gonzalez-Ferez2012}%
  \BibitemOpen
  \bibfield  {author} {\bibinfo {author} {\bibfnamefont {R.}~\bibnamefont {Gonz{\'{a}}lez-F{\'{e}}rez}}\ and\ \bibinfo {author} {\bibfnamefont {C.~P.}\ \bibnamefont {Koch}},\ }\href {https://doi.org/10.1103/PhysRevA.86.063420} {\bibfield  {journal} {\bibinfo  {journal} {Phys. Rev. A}\ }\textbf {\bibinfo {volume} {86}},\ \bibinfo {pages} {063420} (\bibinfo {year} {2012})}\BibitemShut {NoStop}%
\bibitem [{\citenamefont {Crubellier}\ \emph {et~al.}(2015)\citenamefont {Crubellier}, \citenamefont {Gonz{\'{a}}lez-F{\'{e}}rez}, \citenamefont {Koch},\ and\ \citenamefont {Luc-Koenig}}]{Crubellier2015}%
  \BibitemOpen
  \bibfield  {author} {\bibinfo {author} {\bibfnamefont {A.}~\bibnamefont {Crubellier}}, \bibinfo {author} {\bibfnamefont {R.}~\bibnamefont {Gonz{\'{a}}lez-F{\'{e}}rez}}, \bibinfo {author} {\bibfnamefont {C.~P.}\ \bibnamefont {Koch}},\ and\ \bibinfo {author} {\bibfnamefont {E.}~\bibnamefont {Luc-Koenig}},\ }\href {https://doi.org/10.1088/1367-2630/17/4/045020} {\bibfield  {journal} {\bibinfo  {journal} {New J. Phys.}\ }\textbf {\bibinfo {volume} {17}},\ \bibinfo {pages} {045020} (\bibinfo {year} {2015})}\BibitemShut {NoStop}%
\bibitem [{\citenamefont {Shore}(2011)}]{Shore2011}%
  \BibitemOpen
  \bibfield  {author} {\bibinfo {author} {\bibfnamefont {B.}~\bibnamefont {Shore}},\ }\href {https://doi.org/10.1017/CBO9780511675713} {\emph {\bibinfo {title} {Manipulating Quantum Structures Using Laser Pulses}}}\ (\bibinfo  {publisher} {Cambridge University Press},\ \bibinfo {address} {Mexico City},\ \bibinfo {year} {2011})\ p.\ \bibinfo {pages} {586}\BibitemShut {NoStop}%
\bibitem [{\citenamefont {Hioe}(1987)}]{Hioe1987}%
  \BibitemOpen
  \bibfield  {author} {\bibinfo {author} {\bibfnamefont {F.~T.}\ \bibnamefont {Hioe}},\ }\href {https://doi.org/10.1364/josab.4.001327} {\bibfield  {journal} {\bibinfo  {journal} {J. Opt. Soc. Am. B: Opt. Phys.}\ }\textbf {\bibinfo {volume} {4}},\ \bibinfo {pages} {1327} (\bibinfo {year} {1987})}\BibitemShut {NoStop}%
\bibitem [{\citenamefont {Borsalino}\ \emph {et~al.}(2014)\citenamefont {Borsalino}, \citenamefont {Londo{\~{n}}o-Flor{\`{e}}z}, \citenamefont {Vexiau}, \citenamefont {Dulieu}, \citenamefont {Bouloufa-Maafa},\ and\ \citenamefont {Luc-Koenig}}]{Borsalino2014}%
  \BibitemOpen
  \bibfield  {author} {\bibinfo {author} {\bibfnamefont {D.}~\bibnamefont {Borsalino}}, \bibinfo {author} {\bibfnamefont {B.}~\bibnamefont {Londo{\~{n}}o-Flor{\`{e}}z}}, \bibinfo {author} {\bibfnamefont {R.}~\bibnamefont {Vexiau}}, \bibinfo {author} {\bibfnamefont {O.}~\bibnamefont {Dulieu}}, \bibinfo {author} {\bibfnamefont {N.}~\bibnamefont {Bouloufa-Maafa}},\ and\ \bibinfo {author} {\bibfnamefont {E.}~\bibnamefont {Luc-Koenig}},\ }\href {https://doi.org/10.1103/PhysRevA.90.033413} {\bibfield  {journal} {\bibinfo  {journal} {Phys. Rev. A}\ }\textbf {\bibinfo {volume} {90}},\ \bibinfo {pages} {033413} (\bibinfo {year} {2014})}\BibitemShut {NoStop}%
\bibitem [{\citenamefont {Colbert}\ and\ \citenamefont {Miller}(1992)}]{Colbert1992}%
  \BibitemOpen
  \bibfield  {author} {\bibinfo {author} {\bibfnamefont {D.~T.}\ \bibnamefont {Colbert}}\ and\ \bibinfo {author} {\bibfnamefont {W.~H.}\ \bibnamefont {Miller}},\ }\href {https://doi.org/10.1063/1.462100} {\bibfield  {journal} {\bibinfo  {journal} {J. Chem. Phys.}\ }\textbf {\bibinfo {volume} {96}},\ \bibinfo {pages} {1982} (\bibinfo {year} {1992})}\BibitemShut {NoStop}%
\bibitem [{\citenamefont {Forrey}\ \emph {et~al.}(1999)\citenamefont {Forrey}, \citenamefont {Kharchenko}, \citenamefont {Balakrishnan},\ and\ \citenamefont {Dalgarno}}]{Forrey1999}%
  \BibitemOpen
  \bibfield  {author} {\bibinfo {author} {\bibfnamefont {R.~C.}\ \bibnamefont {Forrey}}, \bibinfo {author} {\bibfnamefont {V.}~\bibnamefont {Kharchenko}}, \bibinfo {author} {\bibfnamefont {N.}~\bibnamefont {Balakrishnan}},\ and\ \bibinfo {author} {\bibfnamefont {A.}~\bibnamefont {Dalgarno}},\ }\href {https://doi.org/10.1103/physreva.59.2146} {\bibfield  {journal} {\bibinfo  {journal} {Phys. Rev. A.}\ }\textbf {\bibinfo {volume} {59}},\ \bibinfo {pages} {2146} (\bibinfo {year} {1999})}\BibitemShut {NoStop}%
\bibitem [{\citenamefont {Leibfried}(2012)}]{Leibfried2012}%
  \BibitemOpen
  \bibfield  {author} {\bibinfo {author} {\bibfnamefont {D.}~\bibnamefont {Leibfried}},\ }\href {https://doi.org/10.1088/1367-2630/14/2/023029} {\bibfield  {journal} {\bibinfo  {journal} {New. J. Phys.}\ }\textbf {\bibinfo {volume} {14}},\ \bibinfo {pages} {1} (\bibinfo {year} {2012})}\BibitemShut {NoStop}%
\bibitem [{\citenamefont {Oliveira}\ \emph {et~al.}(2019)\citenamefont {Oliveira}, \citenamefont {Addis}, \citenamefont {Gay}, \citenamefont {Ertel}, \citenamefont {Galimberti},\ and\ \citenamefont {Musgrave}}]{Oliveira2019}%
  \BibitemOpen
  \bibfield  {author} {\bibinfo {author} {\bibfnamefont {P.}~\bibnamefont {Oliveira}}, \bibinfo {author} {\bibfnamefont {S.}~\bibnamefont {Addis}}, \bibinfo {author} {\bibfnamefont {J.}~\bibnamefont {Gay}}, \bibinfo {author} {\bibfnamefont {K.}~\bibnamefont {Ertel}}, \bibinfo {author} {\bibfnamefont {M.}~\bibnamefont {Galimberti}},\ and\ \bibinfo {author} {\bibfnamefont {I.}~\bibnamefont {Musgrave}},\ }\href {https://doi.org/10.1364/OE.27.006607} {\bibfield  {journal} {\bibinfo  {journal} {Opt. Express}\ }\textbf {\bibinfo {volume} {27}},\ \bibinfo {pages} {6607} (\bibinfo {year} {2019})}\BibitemShut {NoStop}%
\bibitem [{\citenamefont {Zou}\ \emph {et~al.}(2016)\citenamefont {Zou}, \citenamefont {Gupta},\ and\ \citenamefont {Caloz}}]{Zou2016}%
  \BibitemOpen
  \bibfield  {author} {\bibinfo {author} {\bibfnamefont {L.}~\bibnamefont {Zou}}, \bibinfo {author} {\bibfnamefont {S.}~\bibnamefont {Gupta}},\ and\ \bibinfo {author} {\bibfnamefont {C.}~\bibnamefont {Caloz}},\ }\href {https://doi.org/10.1109/LMWC.2017.2690880} {\bibfield  {journal} {\bibinfo  {journal} {IEEE Microw. Wirel. Compon. Lett.}\ }\textbf {\bibinfo {volume} {27}} (\bibinfo {year} {2016})}\BibitemShut {NoStop}%
\end{thebibliography}%

\end{document}